\documentclass[%
reprint,
superscriptaddress,
 amsmath,amssymb,
 aps,
prb,
]{revtex4-2}

\usepackage{amsmath}
\usepackage{graphicx}% Include figure files
\usepackage{dcolumn}% Align table columns on decimal point
\usepackage{bm}% bold math
\usepackage{hyperref}% add hypertext capabilities
\usepackage{url}
\usepackage{xcolor}

\hypersetup{colorlinks=true,allcolors=blue}

\newcommand{\Sr}{\texorpdfstring{Sr$_{2}$RuO$_{4}$}{Sr2RuO4}}

\begin{document}

\preprint{APS/123-QED}

\title{Shadowed triplet pairings in Hund's metals with spin-orbit coupling}

\author{Jonathan Clepkens}
\affiliation{Department of Physics and Center for Quantum Materials,\\ University of Toronto, 60 St. George St., Toronto, Ontario, M5S 1A7, Canada}
\author{Austin W. Lindquist}
\affiliation{Department of Physics and Center for Quantum Materials,\\ University of Toronto, 60 St. George St., Toronto, Ontario, M5S 1A7, Canada}
\author{Hae-Young Kee}
 \email{hykee@physics.utoronto.ca}
\affiliation{Department of Physics and Center for Quantum Materials,\\ University of Toronto, 60 St. George St., Toronto, Ontario, M5S 1A7, Canada}
\affiliation{Canadian Institute for Advanced Research, Toronto, Ontario, M5G 1Z8, Canada}


\begin{abstract}
Hund's coupling in multiorbital systems allows for the possibility of even-parity orbital-antisymmetric spin-triplet pairing, which can be stabilized by spin-orbit coupling (SOC). While this pairing expressed in the orbital basis is uniform and spin-triplet, it appears in the band basis as a pseudospin-singlet, with the momentum dependence determined by the SOC and the underlying triplet character remaining in the form of interband pairing active away from the Fermi energy. Here, we examine the role of momentum-dependent SOC in generating nontrivial pairing symmetries, as well as the hidden triplet nature associated with this interorbital pairing, which we dub a {\it ``shadowed triplet"}. Applying this concept to Sr$_{2}$RuO$_{4}$, we first derive several forms of SOC with $d$-wave form-factors from a microscopic model, and subsequently we show that for a range of SOC parameters, a pairing state with $s + id_{xy}$ symmetry can be stabilized. Such a pairing state is distinct from pure spin-singlet and -triplet pairings due to its unique character of pseudospin-energy locking. We discuss experimental probes to differentiate the shadowed triplet pairing from conventional pseudospin-triplet and -singlet pairings.
\end{abstract}

\maketitle

\section{\label{intro} Introduction}
A key feature of superconductivity (SC) is the antisymmetric-wavefunction of the Cooper pair under the exchange of two electrons. This has limited the focus to even-parity spin-singlet and odd-parity spin-triplet pairings in a single Fermi surface (FS) system.
In multiorbital systems, additional possibilities arise due to the orbital degree of freedom, such as even-parity spin-triplet, or odd-parity spin-singlet pairings. For example, it was shown earlier that Hund’s coupling in multiorbital systems allows for an even-parity spin-triplet pairing that is orbitally antisymmetric between two orbitals {\cite{klejnberg1999hund,Dai2008PRL,Puetter2012EPL,Hoshino2015PRL,Hoshino2016PRB, Gingras2019PRL,vafek2017hund}}. The Hund\rq{}s coupling, significant in transition metal systems, acts as an attractive pairing interaction {\cite{klejnberg1999hund}}. Similar to the attractive Hubbard model, the Hund\rq{}s coupling allows a strong local Cooper pair to form between two orbitals with spin-triplet character. 

However, when the electron motion is introduced, i.e., the kinetic term in the Hamiltonian, the pairing is drastically weakened, because the electrons comprising the Cooper pair with momenta ${\bf k}$ and $-{\bf k}$ in different orbitals have different energies, due to the different electronic dispersions of the orbitals. Thus, to stabilize such an orbital antisymmetric pairing, significant orbital degeneracy is required throughout momentum space near the FS {\cite{klejnberg1999hund,Dai2008PRL}}. Furthermore, hybridization between different orbitals can further weaken interorbital pairings, by splitting the degeneracy where it occurs.

The story becomes more interesting when spin-orbit coupling (SOC) is introduced. Note that with SOC, purely spin-singlet and -triplet pairings are no longer well-defined, since the spin and orbital degrees of freedom are coupled. Thus one defines Cooper pairs in the total angular momentum (pseudospin) basis. When the SOC is strong, this forms the basis for SC in the heavy fermion superconductors, such as UPt$_3$, leading to odd-parity pseudospin-triplet SC \cite{Joynt2002RMP,Kallin2016}. When the SOC is intermediate, i.e., comparable to the orbital degeneracy splitting terms, spin and orbital characters vary continuously, and one can define the pairing in either the orbital or band basis. {\it Spin-orbital} mixing then helps to stabilize the interorbital pairing. It was shown that SOC indeed enhances even-parity orbital-antisymmetric spin-triplet pairing \cite{Puetter2012EPL}.  The SOC not only supports the orbital-antisymmetric spin-triplet pairing, but it also transforms pure spin-triplet pairing into ``both" pseudospin-singlet and -triplet pairings in the band basis.

Furthermore, the form of the SOC is not limited to atomic SOC, i.e., $\textbf{L}_{i}\cdot \textbf{S}_{i}$, where $i$ is the site index, and the precise form can determine the momentum dependence of the superconducting state. 
For example, $s$-wave SC proximate to a topological insulator or strong Rashba SOC with broken inversion symmetry leads to an effective $p+ip$ SC, which is odd-parity and a spinless triplet \cite{Fu2008PRL,Fujimoto2008PRB,Fu2009PRB}. For materials with inversion symmetry, there is still the possibility of even-parity momentum-dependent SOC \cite{YigePRB2014} (\textbf{k}-SOC), which has been discussed in the context of the unconventional superconductor, Sr$_{2}$RuO$_{4}$ \cite{Maeno1994Nature, Mackenzie2003RMP, Kallin2012rpp, Mackenzie2017NPJ,Cheung2019PRB,Ramires2019PRB,Suh2019}. Similar to a Rashba-SOC generated $p+ip$ SC, the inclusion of \textbf{k}-SOC in a microscopic model with $s$-wave pairing is reflected in an intraband pairing with the same momentum character as the SOC.

% our motivation
Here we study a microscopic route to \textbf{k}-SOC and how SOC transforms even-parity interorbital spin-triplet SC into pseudospin-singlet and -triplet SC in a Hund's metal, where the multiorbital nature and strong Hund's coupling are crucial. We also illustrate how pseudospin-triplet interband pairing remains, dubbed a {\it ``shadowed triplet"} away from the Fermi energy. While this SC behaves like a singlet in response to low-energy excitations, its hidden identity shows up at finite magnetic fields, and it can be tested when the field strength reaches an appreciable percentage of the superconducting gap size \cite{lindquist2019distinct}. Applying the concept to Sr$_{2}$RuO$_{4}$, we show that $s+id_{xy}$ pairing is stabilized in the $t_{2g}$ orbitals.

% List of sections
The paper is organized as follows. We first introduce the generic Hamiltonian that we will be concerned with throughout in Sec.~\hyperref[Two]{II}. We then consider a simple but general two-orbital model to show how even-parity spin-triplet pairing arises in Sec.~\hyperref[Three]{III}. This includes the stability conditions and how SOC transforms this pairing into a pseudospin-singlet and -triplet in the Bloch band basis. The SOC in the shadowed triplet not only plays an essential role in enhancing the pairing, but it also determines the pairing symmetry. In Sec.~\hyperref[Four]{IV} we investigate microscopic routes to several \textbf{k}-SOC terms with $d$-wave symmetry, which can lead to various $d$-wave pairing symmetries on the FS. In Sec.~\hyperref[Five]{V}, we apply the shadowed triplet pairing scenario to the prominent unconventional superconductor, Sr$_{2}$RuO$_{4}$ \cite{Maeno1994Nature, Mackenzie2003RMP, Kallin2012rpp, Mackenzie2017NPJ}, for which the SOC has been shown to be important \cite{Pavarini2006PRB, Haverkort2008PRL,Rozbicki2011JPCM,Puetter2012EPL,Veenstra2014PRL,Tamai2019PRX}, and we discuss the leading instability towards $s + id_{xy}$ pairing within a three-orbital model for a range of SOC parameters.

\section{\label{Two} General Microscopic Hamiltonian}

We first introduce the generic Hamiltonian that we will be considering throughout. The Hamiltonian consists of three terms. The kinetic term, $H_{0}$, denotes a tight-binding (TB) model, for which we will discuss the precise form in the subsequent sections. The SOC Hamiltonian, $H_{SOC}$, refers to the atomic SOC written in the basis of $t_{2g}$ orbitals and various momentum-dependent terms, also specified later. The interaction, $H_{int}$, has the form of the Kanamori interactions,
\begin{equation}
\begin{aligned}
H_{int} = &\frac{U}{2}\sum_{i,a,\sigma\neq\sigma'}n_{a,i\sigma}n_{a,i\sigma'} + \frac{U'}{2}\sum_{i,a\neq b,\sigma\sigma'}n_{a,i\sigma}n_{b,i\sigma'}\\& + \frac{J_{H}}{2}\sum_{i, a\neq b,\sigma\sigma'}c^{\dagger}_{a,i\sigma}c^{\dagger}_{b,i\sigma'}c_{a,i\sigma'}c_{b,i\sigma} \\&+\frac{J_{H}}{2}\sum_{i, a\neq b,\sigma\neq\sigma'}c^{\dagger}_{a,i\sigma}c^{\dagger}_{a,i\sigma'}c_{b,i\sigma'}c_{b,i\sigma},
\end{aligned}
\end{equation}
where $U$ and $U'$ are the intra- and interorbital Hubbard repulsions, $J_{H}$ is the Hund's coupling and $c_{a,i\sigma}^{\dagger}$ is an electron operator creating an electron at site $i$ in orbital $a$ with spin $\sigma$. Decoupling these interaction terms into even-parity zero-momentum spin-singlet and spin-triplet order parameters\cite{Puetter2012EPL,vafek2017hund,Suh2019,lindquist2019distinct} gives,
\begin{equation} \label{interactions}
\begin{aligned}
&H_{int} = \frac{4U}{N}\sum_{a,\textbf{k}\textbf{k}'}\hat{\Delta}^{s\dagger}_{a,\textbf{k}}\hat{\Delta}^{s}_{a,\textbf{k}'} \\&+ \frac{2(U'-J_{H})}{N}\sum_{\{a\neq b\},\textbf{k}\textbf{k}'}\hat{\mathbf{d}}_{a/b,\textbf{k}}^{\dagger}\cdot\hat{\mathbf{d}}_{a/b,\textbf{k}'} \\&+\frac{4J_{H}}{N}\sum_{a\neq b,\textbf{k}\textbf{k}'}\hat{\Delta}^{s\dagger}_{a,\textbf{k}}\hat{\Delta}^{s}_{b,\textbf{k}'}\\&+ \frac{2(U'+J_{H})}{N}\sum_{a\neq b,\textbf{k}\textbf{k}'}\hat{\Delta}^{s\dagger}_{a/b,\textbf{k}}\hat{\Delta}^{s}_{a/b,\textbf{k}'},
\end{aligned}
\end{equation}
where $N$ is the number of sites, and the spin-triplet and -singlet order parameters are defined as
\begin{equation} \label{OPs}
\begin{aligned}
\hat{\textbf{d}}_{a/b,\textbf{k}} &= \frac{1}{4}\sum_{\sigma\sigma'}[i\sigma^{y}\boldsymbol{\sigma}]_{\sigma\sigma'}\bigl(c_{a,\textbf{k}\sigma}c_{b,-\textbf{k}\sigma'} - c_{b,\textbf{k}\sigma}c_{a,-\textbf{k}\sigma'}\bigr) \\
\hat{\Delta}^{s}_{a/b,\textbf{k}} &= \frac{1}{4}\sum_{\sigma\sigma'}[i\sigma^{y}]_{\sigma\sigma'}\bigl(c_{a,\textbf{k}\sigma}c_{b,-\textbf{k}\sigma'} + c_{b,\textbf{k}\sigma}c_{a,-\textbf{k}\sigma'}\bigr) \\ \hat{\Delta}^{s}_{a,\textbf{k}} &= \frac{1}{4}\sum_{\sigma\sigma'}[i\sigma^{y}]_{\sigma\sigma'}c_{a,\textbf{k}\sigma}c_{a,-\textbf{k}\sigma'},
\end{aligned}
\end{equation}
and $\{a\neq b\}$ represents a sum over the unique pairs of orbital indices. An attractive interorbital spin-triplet, $\hat{\textbf{d}}_{a/b,\textbf{k}}$, channel is present when $J_{H} > U'$, which will be our focus. While this corresponds to a Hund's coupling larger than the typical value of $\sim0.2U$ for Sr$_{2}$RuO$_{4}$ \cite{Tamai2019PRX, mravlje2011PRL}, this type of pairing instability has also been found in several studies beyond mean field (MF) theory without such a requirement\cite{Hoshino2015PRL,Hoshino2016PRB, Gingras2019PRL}. Furthermore, we note that the interorbital, $\hat{\Delta}^{s}_{a/b,\textbf{k}}$, and intraorbital, $\hat{\Delta}^{s}_{a,\textbf{k}}$, spin-singlet order parameters appear with repulsive interactions, but they can be induced by the spin-triplet order parameters through the SOC. Indeed, calculating the spin-singlet order parameters within a MF spin-triplet pairing state shows that they are generally one order of magnitude smaller than the spin-triplet order parameters.

\color{black}

\begin{figure*}[t!]
\includegraphics[width=175mm]{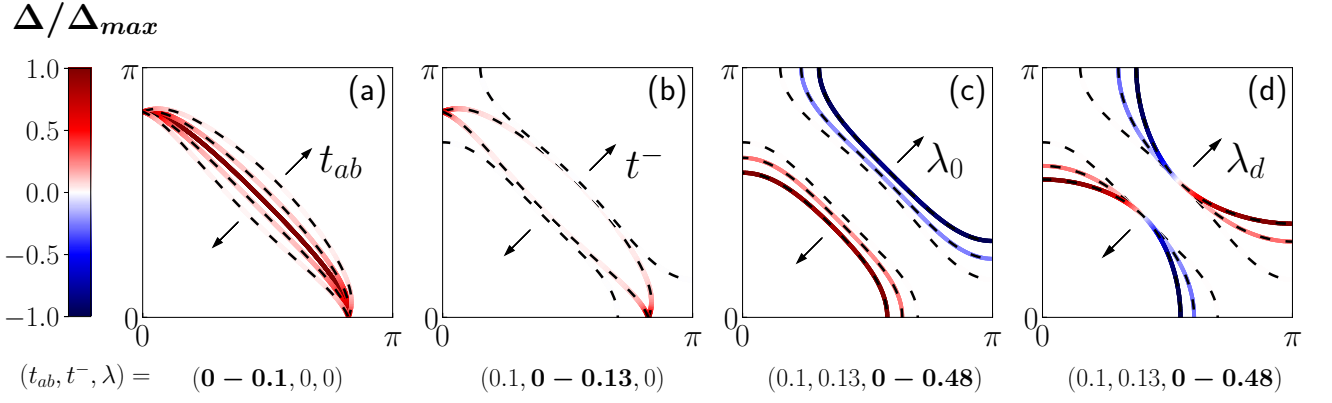}% 
\caption{\label{schematic}Depiction of the gap on the FS in the upper quarter of the first Brillouin zone (BZ) in the two-orbital model, with the arrows indicating the direction the contours move in as the respective parameter is increased and the dashed lines corresponding to the underlying FS. All parameters are given in units of $2t_{1}$; see the main text for other parameter details. (a) The SOC and orbital polarization are set to zero while the strength of the orbital hybridization, $t_{ab}$ in $t_{\textbf{k}} = -4t_{ab}\sin{k_x}\sin{k_y}$, is increased from zero to $0.1$. This shows that the pairing is decreased by the orbital hybridization $t_{ab}$. (b) $t_{ab}$ fixed at $0.1$, and the strength of the orbital polarization, $t^{-}$ in $\xi_{\textbf{k}}^{-} = 2t^{-}(\cos{k_x} - \cos{k_y})$ is increased from zero to $0.13$, resulting in zero gap everywhere. The orbital polarization $t^{-}$ further weakens the pairing. Part (c) shows the revival of the gap for $t_{ab} = 0.1, t^{-} = 0.13$, as the atomic SOC $\lambda_{0}$ is increased from zero to $0.48$ demonstrating that SOC drastically enhances the pairing. Part (d) shows the same as (c) but with the $d$-wave SOC parameter $\lambda_{d}$ in $\lambda_{\textbf{k}} = \lambda_{d}(\cos{k_x} - \cos{k_y})$, instead of $\lambda_{0}$, increased from zero to $0.48$. SOC not only enhances the pairing, but also determines the momentum dependence.}
\end{figure*}

\section{\label{Three} Two-orbital model}
To illustrate how the shadowed triplet pairing occurs, we first consider a two-orbital MF Hamiltonian, $H_{MF}$, consisting of a generic TB model with a specific SOC and an even-parity spin-triplet pairing term. The following model allows for a systematic study of the microscopic components relevant to such pairing in multiorbital systems and is generalized to systems with three orbitals. We use $\Psi_{\textbf{k}}^{\dagger} = (\psi_{\textbf{k}}^{\dagger}, \mathcal{T}\psi_{\textbf{k}}^{T}\mathcal{T}^{-1})$, where $\mathcal{T}$ indicates time-reversal and $\psi_{\textbf{k}}^{\dagger} = (c^{\dagger}_{a,\textbf{k}\uparrow}, c^{\dagger}_{b,\textbf{k}\uparrow}, c^{\dagger}_{a,\textbf{k}\downarrow}, c^{\dagger}_{b,\textbf{k}\downarrow})$ consists of electron operators creating an electron in one of the two orbitals $a,b$ with spin $\sigma=\uparrow,\downarrow$. We also introduce the Pauli matrices plus identity matrix, $\rho_{i},\sigma_{i},\tau_{i}, (i = 0,...3)$, in the particle-hole, spin and orbital spaces respectively, where the direct product between them is implied. $H_{MF}$ is given by,
\begin{equation}
\begin{aligned}
    &H_{MF} = \sum_{\textbf{k}}\Psi_{\textbf{k}}^{\dagger}\bigl(H_{0}(\textbf{k}) + H_\text{SOC}^{z}(\textbf{k}) + H_\text{pair}(\textbf{k})\bigr)\Psi_{\textbf{k}} \\
    &H_{0}(\textbf{k}) = \rho_{3}\bigl( \frac{\xi_{\textbf{k}}^{+}}{2}\sigma_{0}\tau_{0} + \frac{\xi_{\textbf{k}}^{-}}{2}\sigma_{0}\tau_{3}+t_{\textbf{k}}\sigma_{0}\tau_{1} \bigr) \\[5pt]
    &H_\text{SOC}^{z}(\textbf{k}) = -\lambda_{\textbf{k}}\rho_{3}\sigma_{3}\tau_{2} \\[5pt]
    &H_\text{pair} = -d^{z}_{a/b} \rho_{2}\sigma_{3}\tau_{2},
\end{aligned}
\end{equation}
 where the TB contribution is given by the sum of the two orbital dispersions, $\xi_{\textbf{k}}^{+}$, and the difference between them, $\xi_{\textbf{k}}^{-}$, which are defined by $\displaystyle \xi_{\textbf{k}}^{\pm} = \xi_{\textbf{k}}^{a} \pm \xi_{\textbf{k}}^{b}$, as well as the orbital hybridization of the two orbitals, $t_{\textbf{k}}$. Here we consider $H_{SOC}^{z}$, with the form $\lambda_{\textbf{k}}L_{z}S_{z}$, to illustrate the effects of the SOC. Later, we derive several possible \textbf{k}-SOC from a microscopic perspective in Sec.~\hyperref[Four]{IV}, which differ from the $L_{z}S_{z}$ form included here. The pairing Hamiltonian is obtained through a MF decoupling of $H_{int}$ and favors the even-parity interorbital spin-triplet order parameter via $H_{SOC}^{z}$, with the $d$-vector aligned along the $z$-direction, and thus $d^{z}_{a/b}$ is given by
 \begin{equation}
d^{z}_{a/b} \equiv (U'-J_{H})\frac{1}{2N}\sum_{\textbf{k}}\langle{\hat{d}^{z}_{a/b,\textbf{k}}}\rangle.
 \end{equation}

% \begin{figure*}[t!]
% \includegraphics[width=175mm]{fig_1.eps}% 
% \caption{\label{schematic}Depiction of the gap on the FS in the upper quarter of the first Brillouin zone (BZ) in the two-orbital model, with the arrows indicating the direction the contours move in as the respective parameter is increased and the dashed lines corresponding to the underlying FS. All parameters are given in units of $2t_{1}$; see the main text for other parameter details. (a) The SOC and orbital polarization are set to zero while the strength of the orbital hybridization, $t_{ab}$ in $t_{\textbf{k}} = -4t_{ab}\sin{k_x}\sin{k_y}$, is increased from zero to $0.1$. This shows the pairing is decreased by the orbital hybridization $t_{ab}$. (b) $t_{ab}$ fixed at $0.1$, and the strength of the orbital polarization, $t^{-}$ in $\xi_{\textbf{k}}^{-} = 2t^{-}(\cos{k_x} - \cos{k_y})$ is increased from zero to $0.13$, resulting in zero gap everywhere. The orbital polarization $t^{-}$ further weakens the pairing. (c) shows the revival of the gap for $t_{ab} = 0.1, t^{-} = 0.13$, as the atomic SOC $\lambda_{0}$ is increased from zero to $0.48$ demonstrating that SOC drastically enhances the pairing. (d) shows the same as (c) but with the $d$-wave SOC parameter $\lambda_{d}$ in $\lambda_{\textbf{k}} = \lambda_{d}(\cos{k_x} - \cos{k_y})$, instead of $\lambda_{0}$, increased from zero to $0.48$. SOC not only enhances the pairing, but also determines the momentum dependence.}
% \end{figure*}

To understand the stability of the SC state, we consider the relationship between the various components of our model and their effects on the quasi-particle (QP) dispersion. The gap on the FS is shown in Fig.~\ref{schematic} for various cases of $\xi_{\textbf{k}}^{-}$, $t_{\textbf{k}}$ and $\lambda_{\textbf{k}}$. 
Here, $\displaystyle \xi_{\textbf{k}}^{a} = -2t_{1}\cos{k_y} - 2t_{2}\cos{k_x} - \mu$, and for $\xi_{\textbf{k}}^{b}$ we take $x \leftrightarrow y$. The orbitals are coupled through the SOC, for which we take two cases: the atomic SOC denoted by $\lambda_{0}$, and a $d$-wave SOC given by $\lambda_{\textbf{k}} = \lambda_{d}(\cos{k_x} - \cos{k_y})$, as well as the orbital hybridization, which we take as $\displaystyle t_{\textbf{k}} = -4t_{ab}\sin{k_x}\sin{k_y}$. With these dispersions, $\xi_{\textbf{k}}^{-} = 2t^{-}(\cos{k_x} - \cos{k_y})$, where $t^{-} = t_{1} - t_{2}$ and all parameters are given in units of $2t_{1} = 1$. The gap over the FS is shown for four cases: (a) $\lambda=0$, $\xi_{\textbf{k}}^{-}=0$, as $t_{ab}$ is increased from zero to $0.1$; (b) keeping $\lambda=0$, with $t_{ab} = 0.1$ as $\xi_{\textbf{k}}^{-}$ is increased by tuning $t^{-}$ from zero to $0.13$; (c) both $t_{ab}$ and $t^{-}$ are kept the same as $\lambda_{0}$ is increased from zero to 0.48; and (d) the same as (c) but with the $d$-wave SOC, instead of $\lambda_{0}$,  increased by tuning $\lambda_{d}$ from zero to 0.48. 

When the orbital dispersions are completely degenerate, we see that the gap is non-zero everywhere over the two identical bands, as shown by the middle contour in Fig.~\hyperref[schematic]{1(a)}. As the strength of the hybridization, $t_{\textbf{k}}$, is increased from zero, the energy separation of the two orbital dispersions increases wherever $t_{\textbf{k}}\neq0$ and the gap arising from interband pairing disappears on the FS, except where $t_{\textbf{k}}$ vanishes along the $k_{x/y} = 0$ lines, where the two orbital dispersions remain degenerate. Starting from there with zero interband pairing over most of the FS, Fig.~\hyperref[schematic]{1(b)} demonstrates the effect of $\xi_{\textbf{k}}^{-}$ in further reducing the interband pairing to zero everywhere on the FS, due to the absence of phase space for zero-momentum pairing. Keeping the same parameters used in (b), Fig.~\hyperref[schematic]{1(c)} reveals how the SOC revives the SC state by allowing for an intraband pairing on the FS. As the SOC is increased from zero, the intraband gap becomes non-zero over the entire FS. Additionally, the sign of the gap function is opposite on the two bands, matching the $\Delta^{s}(\textbf{k})\tilde{\tau}_{3}$ dependence, which we will show below in Eq.~(\ref{bandham}) and Eq.~(\ref{bandham2}), but uniform on each band due to the lack of momentum dependence of the atomic SOC. In contrast, Fig.~\hyperref[schematic]{1(d)} displays the $d$-wave dependence of the gap arising from the $d$-wave SOC. Thus with the introduction of $t_{\textbf{k}}$, $\xi_{\textbf{k}}^{-}$ and $\lambda_{\textbf{k}}$, the pairing on the FS is transformed from an interband spin-triplet to a purely intraband pseudospin-singlet with the same momentum dependence as the associated SOC, while the pseudospin-triplet is active away from the Fermi energy.

The above results can also be understood via the commutation relations between the order parameter and other terms. The pair-breaking effects are revealed by the commuting behavior with the pairing term \cite{fischer2013NJP,Ramires2016PRB,Ramires2018PRB}. Conversely, the SOC anti-commutes with the pairing term and generally enhances the pairing state.
The anti-commutators are given by 
\begin{equation}
\begin{aligned}    
    \{\frac{\xi_{\textbf{k}}^{-}}{2}\rho_{3}\sigma_{0}\tau_{3},   H_\text{pair}\} &= \xi_{\textbf{k}}^{-}d^{z}_{a/b}\rho_{1}\sigma_{3}\tau_{1} \\
    \{t_{\textbf{k}}\rho_{3}\sigma_{0}\tau_{1}, H_\text{pair}\} &= -2t_{\textbf{k}}d^{z}_{a/b}\rho_{1}\sigma_{3}\tau_{3} \\
    \{-\lambda_{\textbf{k}}\rho_{3}\sigma_{3}\tau_{2},H_\text{pair}\} &= 0.
\end{aligned}
\end{equation}
These effects are reflected in the QP dispersion, given by
\begin{equation}
\begin{aligned}
    E_{\textbf{k}} &= \pm\frac{1}{2}\biggl[\xi_{\textbf{k}}^{-2} + \xi_{\textbf{k}}^{+2} + 4\bigl[t_{\textbf{k}}^2 + (d^{z}_{a/b})^2 + \lambda_{\textbf{k}}^2\bigr] \\ &\pm 2\sqrt{\xi_{\textbf{k}}^{+2}\bigl[\xi_{\textbf{k}}^{-2} + 4t_{\textbf{k}}^2 + 4\lambda_{\textbf{k}}^2\bigr] + 4(d^{z}_{a/b})^2\bigl[\xi_{\textbf{k}}^{-2} + 4t_{\textbf{k}}^2\bigr]} \biggr]^{\frac{1}{2}}.
\end{aligned}
\end{equation}
The general equation of the FS is $\displaystyle \xi_{\textbf{k}}^{-2} = \xi_{\textbf{k}}^{+2} - 4(t_{\textbf{k}}^2 + \lambda_{\textbf{k}}^2)$, from which it can be seen that if $\lambda_{\textbf{k}}$ and  $t_{\textbf{k}} = 0$, and the orbitals are degenerate, i.e.,~$\xi_{\textbf{k}}^{-} = 0$, we recover the conventional BCS result for the gap energy on the FS: $\pm d^{z}_{a/b}$. In the general case where these terms are non-zero, assuming $d^{z}_{a/b}$ is small, the QP gap is
\begin{equation}
\text{QP gap}_{\text{on FS}} \approx \pm\sqrt{\frac{(d^{z}_{a/b})^4+16\lambda_{\textbf{k}}^2(d^{z}_{a/b})^2}{4(\xi_{\textbf{k}}^{-2} + 4(t_{\textbf{k}}^2 + \lambda_{\textbf{k}}^2))}}.
\end{equation} 
From this, it is clear that increasing $\xi_\textbf{k}^-$ and $t_\textbf{k}$ decreases the overall gap size. 
The detrimental effects on the gap size, signified by the commuting nature of both $t_{\textbf{k}}$ and $\xi_{\textbf{k}}^{-}$ with the pairing Hamiltonian, are a result of shifting apart in energy the bands being paired, resulting in the gap moving away from the FS.
However, turning on the SOC significantly enhances the gap size \cite{Puetter2012EPL}, as shown in Fig.~\hyperref[schematic]{1(c)}, and as the SOC strength becomes larger than $\xi_{\textbf{k}}^{-}$ and $t_{\textbf{k}}$, the gap can be restored to the order of $d^z_{a/b}$. Furthermore, if $\lambda_{\textbf{k}}$ has a $d$-wave form factor such as ($\cos{k_x}-\cos{k_y})$, the gap reflects the exact same $d$-wave symmetry, as shown in Fig.~\hyperref[schematic]{1(d)}. The enhancement of the SC state is accomplished by providing a non-zero intraband pseudospin-singlet pairing on the FS, as we show below in the band basis.

With the aim to further understand the nature of the SC state, we study how the pairing transforms to the Bloch band basis, labeled by band indices $\alpha, \beta$ and pseudospin $s = (+,-)$. The transformation is given by
\begin{equation}
\begin{aligned}
\begin{pmatrix}
   c_{a,\textbf{k}\sigma} \\[6pt]
   c_{b,\textbf{k}\sigma} \\
 \end{pmatrix}
 = \begin{pmatrix} 
    \frac{\eta_{\sigma}+1}{2}f_{\textbf{k}}-\frac{\eta_{\sigma}-1}{2}f_{\textbf{k}}^{*} & -g_{\textbf{k}}  \\[6pt]
  g_{\textbf{k}} & \frac{\eta_{\sigma}+1}{2}f_{\textbf{k}}^{*}-\frac{\eta_{\sigma}-1}{2}f_{\textbf{k}}
\end{pmatrix}
\begin{pmatrix}
   \alpha_{\textbf{k},s} \\[6pt]
   \beta_{\textbf{k},s} \\
 \end{pmatrix}
\end{aligned}
\end{equation}
where $\eta_{\sigma}=+1(-1)$ for $\sigma=\uparrow(\downarrow)$ and $s=+(-)$. The coefficients of the transformation are given by $f_{\textbf{k}} = -\frac{\gamma_{\textbf{k}}}{|\gamma_{\textbf{k}}|}\sqrt{\frac{1}{2}(1 + \frac{\xi_{\textbf{k}}^{-}}{\sqrt{\xi_{\textbf{k}}^{-2} + 4|\gamma_{\textbf{k}}|^2 }})}$ and $g_{\textbf{k}} = -\sqrt{\frac{1}{2}(1-\frac{\xi_{\textbf{k}}^{-}}{\sqrt{\xi_{\textbf{k}}^{-2} + 4|\gamma_{\textbf{k}}|^2 }})}$, where $f_{\textbf{k}}$ is chosen to be complex with the same phase as $\gamma_{\textbf{k}} = t_{\textbf{k}} + i\lambda_{\textbf{k}}$, $g_{\textbf{k}}$ is real, and $|f_{\textbf{k}}|^2 + g_{\textbf{k}}^2 = 1$. Applying this transformation on $H_\text{pair}$ results in the pairing in the band basis, where we have also defined the Pauli matrices $\tilde{\rho}_{i}, \tilde{\sigma}_{i}, \tilde{\tau}_{i}$ in the Nambu, pseudospin, and band spaces with the basis $\Phi_{\textbf{k}}^{\dagger} = (\phi_{\textbf{k}}^{\dagger},\mathcal{T}\phi_{\textbf{k}}^{T}\mathcal{T}^{-1})$ and $\phi_{\textbf{k}}^{\dagger} = (\alpha_{\textbf{k}+}^{\dagger},\beta_{\textbf{k}+}^{\dagger},\alpha_{\textbf{k}-}^{\dagger},\beta_{\textbf{k}-}^{\dagger})$. We obtain,
\begin{equation} \label{bandham}
\begin{aligned}
    \tilde{H}_\text{pair}(\textbf{k}) &= \tilde{\rho}_{2}\tilde{\sigma}_{0}\bigl[-\Delta^{s}(\textbf{k})\tilde{\tau}_{3} - \Delta^{s}_{\alpha/\beta}(\textbf{k})\tilde{\tau}_{1}\bigr]\\  &-d^{z}_{\alpha/\beta}(\textbf{k})\tilde{\rho}_{2}\tilde{\sigma}_{3}\tilde{\tau}_{2},
\end{aligned}
\end{equation}
\\
where $\Delta^{s}(\textbf{k})$ and $\Delta^{s}_{\alpha/\beta}(\textbf{k})$ denote pseudospin-singlet intraband and interband pairings respectively, and $d^{z}_{\alpha/\beta}(\textbf{k})$ is a pseudospin-triplet interband pairing. The intraband and interband nature of these pairings becomes more apparent from the operator form,
\begin{widetext}
\begin{equation} \label{bandham2}
\begin{aligned}
\tilde{H}_\text{pair}(\textbf{k}) = i\Delta^{s}(\textbf{k})\bigl[(\alpha_{\textbf{k},+}^{\dagger}\alpha_{-\textbf{k},-}^{\dagger} -\alpha_{\textbf{k},-}^{\dagger}\alpha_{-\textbf{k},+}^{\dagger}) &- (\beta_{\textbf{k},+}^{\dagger}\beta_{-\textbf{k},-}^{\dagger} -\beta_{\textbf{k},-}^{\dagger}\beta_{-\textbf{k},+}^{\dagger})\bigr] \\
+ i\Delta^{s}_{\alpha/\beta}(\textbf{\textbf{k}})\bigl[(\alpha^{\dagger}_{\textbf{k},+}\beta^{\dagger}_{-\textbf{k},-} - \alpha^{\dagger}_{\textbf{k},-}\beta^{\dagger}_{-\textbf{k},+}) &+ (\beta^{\dagger}_{\textbf{k},+}\alpha^{\dagger}_{-\textbf{k},-} - \beta^{\dagger}_{\textbf{k},-}\alpha^{\dagger}_{-\textbf{k},+})\bigr] \\
+d^{z}_{\alpha/\beta}(\textbf{\textbf{k}})\bigl[(\alpha^{\dagger}_{\textbf{k},+}\beta^{\dagger}_{-\textbf{k},-} + \alpha^{\dagger}_{\textbf{k},-}\beta^{\dagger}_{-\textbf{k},+}) &- (\beta^{\dagger}_{\textbf{k},+}\alpha^{\dagger}_{-\textbf{k},-} + \beta^{\dagger}_{\textbf{k},-}\alpha^{\dagger}_{-\textbf{k},+})\bigr].
\end{aligned}
\end{equation}
\end{widetext}
The above equation shows the sign change in the intraband pairing between the two bands, as displayed in Fig.~\hyperref[schematic]{1(c)} and \hyperref[schematic]{1(d)}. This relative sign between the bands is similar to the $s^{+-}$ gap structure \cite{vafek2017hund}, although it should be noted that here for simplicity we have ignored the SOC-induced intraorbital singlets \cite{Puetter2012EPL, vafek2017hund, lindquist2019distinct, Huang2019PRB} that would add to the gap on each band. For small $(J_{H}-U')$ they are small compared to $d^{z}_{a/b}$ and they do not affect the conclusions of this section. The pairings in the band basis have the following form:
\begin{equation}
\begin{aligned}
    \Delta^{s}(\textbf{k}) &= -2d^{z}_{a/b}\text{Im}(f_{\textbf{k}})g_{\textbf{k}} =  -\frac{2d^{z}_{a/b}\lambda_{\textbf{k}}}{\sqrt{\xi_{\textbf{k}}^{-2} + 4(t_{\textbf{k}}^2 + \lambda_{\textbf{k}}^2)}} \\[6pt]
    \Delta^{s}_{\alpha/\beta}(\textbf{k}) &= -d^{z}_{a/b}\text{Im}(f_{\textbf{k}}^2) =  -2d^{z}_{a/b}|f_{\textbf{k}}|^2\frac{t_{\textbf{k}}\lambda_{\textbf{k}}}{t_{\textbf{k}}^2 + \lambda_{\textbf{k}}^2} \\[6pt]
    d^{z}_{\alpha/\beta}(\textbf{k}) &= d^{z}_{a/b}[g_{\textbf{k}}^2 + \text{Re}(f_{\textbf{k}}^2)] = d^{z}_{a/b}\bigl(g_{\textbf{k}}^2 + |f_{\textbf{k}}|^2\frac{t_{\textbf{k}}^2 - \lambda_{\textbf{k}}^2}{t_{\textbf{k}}^2 + \lambda_{\textbf{k}}^2}\bigr).
\end{aligned}
\end{equation}
While the orbital order parameter is $s$-wave and contains no explicit momentum dependence, transforming to the band basis generates potentially complex momentum dependence from SOC and orbital hybridization. The orbital spin-triplet order parameter carries over to the interband pseudospin-triplet, which, in the limit of zero SOC, becomes equal to $d^{z}_{a/b}$, while both the intraband and interband pseudospin-singlets vanish. The interband pseudospin-singlet pairing also vanishes for $t_{\textbf{k}}=0$. However, an important feature for $\lambda_{\textbf{k}}\neq0$ is the presence of the intraband pseudospin-singlet pairing, $\Delta^{s}(\textbf{k})$, which acquires the same symmetry dependence as a function of momentum as $\lambda_{\textbf{k}}$, as shown in Fig.~\hyperref[schematic]{1(d)}. While the interband pseudospin-triplet is a signature of the fundamental interorbital spin-triplet order parameter, it is the intraband pseudospin-singlet that leads to a weak-coupling instability. This is because the interband pairing contribution to the gap will generally be negligible on the FS, such that the gap is given by $|\Delta^{s}|$. Considering the QP dispersion in terms of the band pairings,
\begin{equation}
\begin{aligned}
    &E_{\textbf{k}} = \pm\biggl[\frac{(\xi_{\textbf{k}}^{\alpha})^2 + (\xi_{\textbf{k}}^{\beta})^2}{2} + (\Delta^{s})^2 + (\Delta^{s}_{\alpha/\beta})^2 + (d^{z}_{\alpha/\beta})^2 \\ &\pm \sqrt{\frac{\bigl[(\xi_{\textbf{k}}^{\alpha})^2 - (\xi_{\textbf{k}}^{\beta})^2\bigr]^2}{4} + (\xi_{\textbf{k}}^{\alpha}- \xi_{\textbf{k}}^{\beta})^2\bigl[(\Delta^{s}_{\alpha/\beta})^2 + (d^{z}_{\alpha/\beta})^2\bigr]}\biggr]^{\frac{1}{2}},
\end{aligned}
\end{equation}
and evaluating this on either the $\alpha$ or $\beta$ FS, for only $\Delta^s$ intraband pairing, we obtain the gap energy $\pm |\Delta^{s}|$. For either only $\Delta^s_{\alpha/\beta}$ or $d^z_{\alpha/\beta}$ interband pairing, a gap of $\pm\Delta^s_{\alpha/\beta}$ or $\pm d^z_{\alpha/\beta}$ forms where $\xi_\textbf{k}^\alpha = -\xi_\textbf{k}^\beta$. This corresponds to an energy gap on the FS only where $\xi_\textbf{k}^\alpha = \xi_\textbf{k}^\beta = 0$, which is not a generic feature but rather requires fine-tuning to achieve, otherwise the gap formed will be away from the Fermi energy.

While the model introduced in this section is simple, its generality gives insight into the role of the orbital degeneracy, hybridization, and SOC for multiorbital systems in dictating the stability of the even-parity spin-triplet SC state. It is straightforward to extend to three-orbital descriptions. Furthermore, we have seen that in the band basis, the intraband pairing on the FS takes on the momentum dependence of the SOC, allowing for a rich collection of pairing symmetries unexpected from the original $s$-wave order parameter and in contrast to other forms of momentum-dependent SC that arise from nonlocal interactions. However, the possible pairing symmetries will depend on the forms of \textbf{k}-SOC that can be obtained from microscopic considerations. Therefore, we now turn to a study of how various forms of \textbf{k}-SOC can arise microscopically.

\section{\label{Four} Microscopic route to momentum-dependent SOC}
Here, we take as a specific microscopic example the layered perovskite Sr$_{2}$RuO$_{4}$, which has the tetragonal space group I4/{\it mmm} and point group $D_{4h}$, for which the Ru 4$d$ $t_{2g}$ orbitals are the relevant low-energy degrees of freedom. With this, we study how the various forms of \textbf{k}-SOC with different $d$-wave form factors such as (a) an in-plane $d_{xy}$ SOC in the B$_{2g}$ representation, (b) in-plane $d_{x^2-y^2}$ SOC in the B$_{1g}$ representation, and (c) interlayer $\{d_{xz},d_{yz}\}$ SOC in the E$_{g}$ representation can arise microscopically, going beyond a purely symmetry-based analysis.

\subsection{In-plane \texorpdfstring{B\textsubscript{2g}}{B2g}}

We begin by discussing the in-plane \textbf{k}-SOC in the B$_{2g}$ representation, which contains ($\sin{k_x}\sin{k_y}$) momentum dependence. This SOC is important since based on symmetry, there will already be a finite $H_{SOC}^{B_{2g}}$ in the presence of the orbital hybridization and atomic SOC. Since the orbital hybridization for the system considered here will appear between the $d_{xz}$ and $d_{yz}$ orbitals as $\displaystyle H_{ab} = -4t_{ab}\sum_{\textbf{k}\sigma}\sin{k_x}\sin{k_y}\bigl(c^{\dagger}_{yz,\textbf{k}\sigma}c_{xz,\textbf{k}\sigma} + \text{H.c.}\bigr)$, and transforms under the B$_{2g}$ representation, there is a cubic coupling term in the free energy between the atomic SOC ($\lambda$), transforming as A$_{1g}$, orbital hybridization and B$_{2g}$ SOC. This symmetry-allowed coupling $\displaystyle \sim\langle{H_{SOC}^{B_{2g}}}\rangle\langle{H_{SOC}^{A_{1g}}}\rangle\langle{H_{ab}}\rangle$, where $H_{SOC}^{A_{1g}}$ denotes the atomic SOC, ensures the presence of a non-zero B$_{2g}$ SOC in the presence of both orbital hybridization and atomic SOC.

\begin{figure}[!t]
\includegraphics[width=86.5mm]{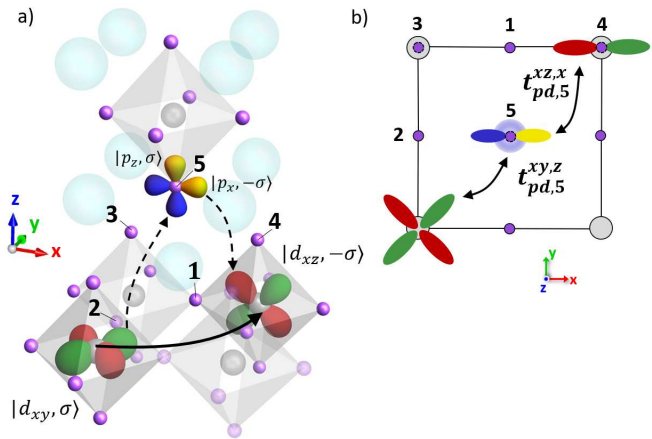}% 
\caption{\label{b2g}Hopping channels generating the in-plane B$_{2g}$ $d$-wave \textbf{k}-SOC for the $d_{xz}$ and $d_{xy}$ orbitals (green and red), which consists of an effective spin-flip hopping between next-nearest neighbor sites. The alternative process involving the $d_{yz}$ orbital is related by a $C_{4}$ rotation and only the interlayer process is shown in detail for clarity. (a) The intermediate oxygen sites for the contributing hopping channels are indicated by the numbering (1-5), with the relevant $p$ orbitals (yellow and blue) and the intermediate hopping amplitudes shown by dashed lines only for the fifth channel. The other channels involve the same $p$-orbitals at sites 1-4. The top layer of the unit cell is also removed for clarity. (b) Schematic top view of the hopping process, where the bottom lobe of the $p_{z}$ orbital is shown.}
\end{figure}

Furthermore, the in-plane \textbf{k}-SOC in the B$_{2g}$ representation arises through several hopping channels, all utilizing intermediate $p$-orbitals, but different oxygen sites denoted by 1-5 in Fig.~\ref{b2g}, including only nearby sites. As most of these channels occur in a single layer of Ru-O octahedra, this SOC should be expected to be the leading contribution beyond the atomic SOC, since we will see that the other two \textbf{k}-SOC always require hopping to additional layers. Such an in-plane \textbf{k}-SOC can be obtained through perturbation theory by considering hopping between next-nearest neighbor Ru atoms through the various channels, via the oxygen sites as intermediate states, and including the oxygen $p$-orbitals' atomic SOC. This hopping process results in an electron hopping from the $d_{xy}$ orbital with spin $\sigma$ to either a $d_{xz}$ or $d_{yz}$ orbital with spin $-\sigma$, where the former case is shown in Fig.~\ref{b2g} by the black solid line. The contributing channels are indicated by the numbering of the intermediate oxygen sites, and only path 5 is shown explicitly for clarity. The effective SOC Hamiltonian involving the next-nearest neighbor Ru sites is obtained by
\begin{equation} \label{effective}
    H_{SOC}^{B_{2g}} = \sum_{p_{\pm}} \frac{H^{0}|p_{\pm}\rangle\langle{p_{\pm}}|H^{0}}{E_{d}-E_{p_{\pm}}},
\end{equation}
where $H^{0}$ denotes the hopping Hamiltonian involving both $d$ and $p$ orbitals. The sum runs over the intermediate oxygen states for all channels, which are eigenstates of the oxygen SOC $|j,m_{j}, \mathbf{r}\rangle$, with $\mathbf{r}$ the position of the oxygen site, and we consider up to the 2nd order of the perturbation theory for this process. 

For instance, considering a hopping process between a $d_{xy}$ state with spin-$\uparrow$ and a spin-$\downarrow$ $d_{xz}$ state via the apical oxygen site shown, we have $|p_{+}\rangle = -\frac{1}{\sqrt{6}}(|p_{x},\downarrow\rangle + i|p_{y},\downarrow\rangle) + \frac{\sqrt{2}}{\sqrt{3}}|p_{z},\uparrow\rangle$ and $|p_{-}\rangle = -\frac{1}{\sqrt{3}}(|p_{x},\downarrow\rangle + i|p_{y},\downarrow\rangle) - \frac{1}{\sqrt{3}}|p_{z},\uparrow\rangle$. The energy denominator is given for $|p_{+}\rangle$ and $|p_{-}\rangle$ by $E_{pd} + \frac{\lambda_{p}}{2}$ and $E_{pd} - \lambda_{p}$ respectively, where $E_{pd}$ is the difference in the on-site atomic potentials, and $\lambda_{p}$ is the oxygen SOC constant. Considering the hopping amplitude between the $d_{xy}$ orbital with spin $\sigma$ at site $\textbf{R}$ denoted by $|xy,\sigma,\textbf{R}\rangle$ and the opposite spin state of the $d_{xz}$ orbital at site $\textbf{R}'$, $|xz,-\sigma,\textbf{R}'\rangle$, we obtain,
\begin{equation} \label{b2g_amp}
\begin{aligned}
    &\langle{xz,-\sigma,\textbf{R}'}|H_{SOC}^{B_{2g}}|xy,\sigma,\textbf{R}\rangle =\\[6pt] & \eta_{\sigma}\frac{\lambda_{p}}{(E_{pd} + \frac{\lambda_{p}}{2})(E_{pd} - \lambda_{p})}\sum_{i}t_{pd,i}^{a_{i}}t_{pd,i}^{b_{i}},
    \end{aligned}
\end{equation}
where the sum is over the contributing channels involving the different oxygen sites indicated in Fig.~\ref{b2g} and $a_{i}, b_{i}$ refer to the orbitals in the intermediate hopping amplitudes. All of these channels contain the hopping amplitudes between either the $d_{xy}/p_{x}$ and $d_{xz}/p_{z}$, or $d_{xy}/p_{z}$ and $d_{xz}/p_{x}$ orbitals. From Fig.~\ref{b2g} it can be seen that the overall sign dependence of the hopping channel will match that of the $d_{xy}$ orbital, since $p_{x}$ and $d_{xz}$ change sign identically under the $yz$ and $xz$ mirror planes, while $p_{z}$ is even under them, leading to the $\sin{k_x}\sin{k_y}$ dependence. Furthermore, the presence of $\eta_{\sigma}$ in Eq.~(\ref{b2g_amp}) gives rise to $\sigma^{y}$ spin-dependence.

Taking into account the equivalent process between the $d_{xy}$ and $d_{yz}$ orbitals, which is related by a $C_{4}$ rotation, we obtain
\begin{equation} \label{B2gEqn_1}
\begin{aligned}
    H_{SOC}^{B_{2g}} &= 4i\lambda^{B_{2g}}\sum_{\textbf{k}\sigma\sigma'}\sin{k_x}\sin{k_y} \sigma^{y}_{\sigma\sigma'}c^{\dagger}_{xz,\textbf{k}\sigma}c_{xy,\textbf{k}\sigma'} \\&-4i\lambda^{B_{2g}}\sum_{\textbf{k}\sigma\sigma'}\sin{k_x}\sin{k_y} \sigma^{x}_{\sigma\sigma'}c^{\dagger}_{yz,\textbf{k}\sigma}c_{xy,\textbf{k}\sigma'} + \text{H.c.},
\end{aligned}
\end{equation}
where $\lambda^{B_{2g}} = \frac{\lambda_{p}}{(E_{pd} + \frac{\lambda_{p}}{2})(E_{pd} - \lambda_{p})}\sum_{i}t_{pd,i}^{a_{i}}t_{pd,i}^{b_{i}}$
and $i=1,..,5$ indicates the different $p$-orbital sites involved in Fig.~\ref{b2g}.

Quantifying the value of $\lambda^{B_{2g}}$ in Sr$_{2}$RuO$_{4}$ will require further studies to accurately include the additional effects of the coupling between other microscopic parameters such as the Ru on-site SOC and orbital hybridization, as discussed above. Correlation effects have also been shown to be crucial for the size of SOC, which is enhanced from the local-density approximation SOC values \cite{Tamai2019PRX}. The current work is to show a microscopic route to generating \textbf{k}-SOC within a perturbation theory approach, beyond a purely symmetry-based perspective. These considerations also apply to the other \textbf{k}-SOC processes, to which we now turn.

\subsection{In-plane \texorpdfstring{B\textsubscript{1g}}{B1g}}
\begin{figure}[!t]
\includegraphics[width=86mm]{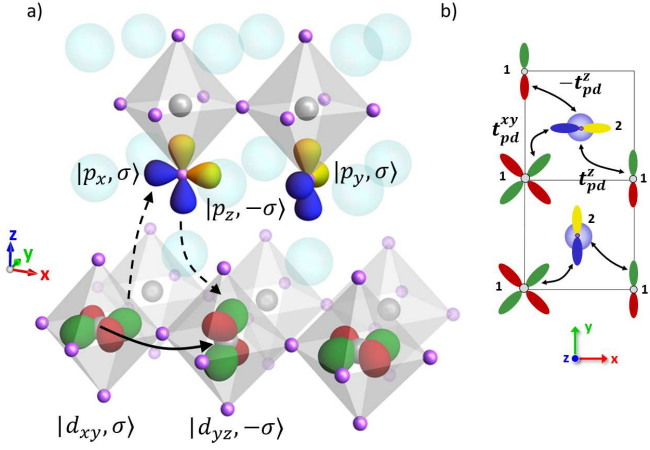}% 
\caption{\label{dwave}Hopping processes generating the in-plane B$_{1g}$ $\textbf{k}$-SOC, which consists of an effective spin-flip hopping between nearest neighbor sites indicated by the solid line. (a) The relevant $d$ and $p$ orbitals shown in the three-dimensional (3D) structure and the dashed lines indicating the intermediate hopping processes. The process involving the $d_{yz}$ orbital is shown as an example, with the alternative process involving the $d_{xz}$ orbital related by a $C_{4}$ rotation. The $p_{x}$ and $p_{y}$ orbitals are separated for clarity. (b) Schematic picture of the hopping process from a top view where only the lobe of the $p_{z}$ wavefunction closest to the plane of the hopping is shown and the numbers indicate whether the orbital is on the apical oxygen site (2) or bottom (1) layer. The $p_{x}$ and $p_{y}$ orbitals are also shown in two separate squares for clarity.}
\end{figure}

We now consider a microscopic route to obtaining an in-plane \textbf{k}-SOC in the B$_{1g}$ representation, which has a $d_{x^2-y^2}$ form factor. This requires a different layer of Ru-O octahedra, but the same procedure. An example of this process is shown in Fig.~\hyperref[dwave]{3(a)}, for which the hopping between the $d_{xy}$ orbital with a spin $\sigma$ state and the opposite spin state of the $d_{yz}$ orbital is indicated, where the overlap of the $d_{xy}$ orbital with both the $p_{x}$ and $p_{y}$ orbitals is shown. The hopping amplitudes are represented schematically in Fig.~\hyperref[dwave]{3(b)} for both the $+\hat{x}$ or $+\hat{y}$ directions, where the $p_{z}$ lobe closest to the plane containing the effective hopping is shown, and the $p_{x}$/$p_{y}$ orbitals are in separate squares for clarity. The numbering in Fig.~\hyperref[dwave]{3(b)} indicates whether the orbital is on the apical oxygen site (2) or the bottom (1) layer.  We denote the hopping between the $d_{xy}$ orbital and $p_{x}$($p_{y}$) as $t_{pd}^{xy,x}(t_{pd}^{xy,y})$, with both having magnitude $t_{pd}^{xy}$ due to the $C_{4}$ rotational symmetry. The hopping between either of the $d_{yz}/d_{xz}$ orbitals and $p_{z}$ is denoted by $t_{pd}^{z}$, where the two are also of equal magnitude.

Evaluating the sum over the possible intermediate states for a given apical oxygen site, we obtain
\begin{equation} \label{amplitude}
\begin{aligned}
&\langle{yz,-\sigma,\textbf{R}'}|H_{SOC}^{B_{1g}}|xy,\sigma,\textbf{R}\rangle = \\[6pt] &-\frac{\lambda_{p}}{2(E_{pd} + \frac{\lambda_{p}}{2})(E_{pd} - \lambda_{p})}\sum_{\textbf{r}}t_{pd}^{z}\biggl[ \eta_{\sigma}t_{pd}^{xy,x} + it_{pd}^{xy,y}\biggr].
\end{aligned}
\end{equation}
Due to the mirror symmetry about the $xz$ plane, where the apical oxygen above the plane in the $+\hat{x}+\hat{y}$ direction is shown in Fig.~\ref{dwave}, the imaginary term cancels after summing over the two oxygen sites at $\pm \hat{y}$, since $t_{pd}^{z}t_{pd}^{xy,y} \rightarrow - t_{pd}^{z}t_{pd}^{xy,y}$, while the real term is invariant. Summing over both $x$- and $y$-direction paths, we have,
\begin{equation}
\begin{aligned}
H_{SOC}^{B_{1g}} &= 2i\lambda^{B_{1g}}\sum_{\textbf{k}\sigma\sigma'}\sigma^{y}_{\sigma\sigma'}(\cos{k_x} - \cos{k_y})c^{\dagger}_{yz,\textbf{k}\sigma}c_{xy,\textbf{k}\sigma'} \\ &+2i\lambda^{B_{1g}}\sum_{\textbf{k}\sigma\sigma'}\sigma^{x}_{\sigma\sigma'}(\cos{k_x} - \cos{k_y})c^{\dagger}_{xz,\textbf{k}\sigma}c_{xy,\textbf{k}\sigma'}\\& + \text{H.c.},
\end{aligned}
\end{equation}
where $\lambda^{B_{1g}} = 2\frac{t_{pd}^{xy}t_{pd}^{z}\lambda_{p}}{(E_{pd} + \frac{\lambda_{p}}{2})(E_{pd} - \lambda_{p})}$ and it can be noted that this Hamiltonian is similar to $\displaystyle \lambda(L_{x}S_{x} - L_{y}S_{y})$, but with $\lambda$ replaced with a $d_{x^2-y^2}$ form factor.

\subsection{Interlayer \texorpdfstring{E\textsubscript{g}}{Eg}  }
\begin{figure}[!t]
\includegraphics[width=86mm]{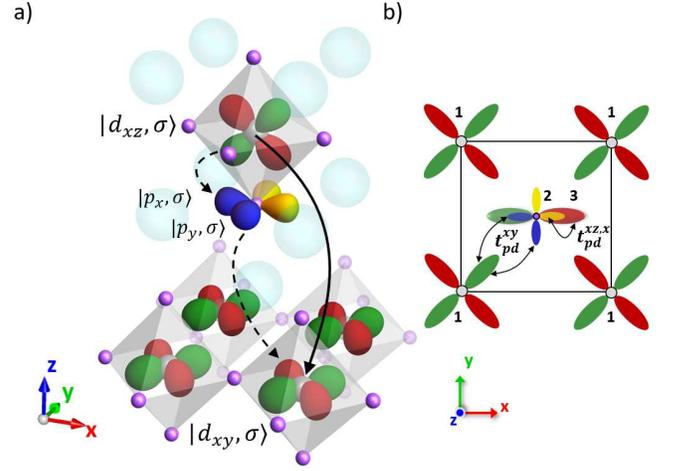}% 
\caption{\label{interlayer}Hopping process generating the E$_{g}$ $\textbf{k}$-SOC and interlayer hopping. (a) The relevant $d$ and $p$ orbitals shown in the 3D structure, where the dashed lines indicate the intermediate hopping processes. Only the process involving the $d_{xz}$ orbital hopping in the $-\hat{z}$ direction is shown for clarity. (b) A schematic picture of the hopping process from a top view where the bottom lobes of the $d_{xz}$ orbital are shown on top of the $p_{x}$ orbital and the numbers indicate whether the orbital is on the top (3), bottom (1) layer or the apical oxygen site (2).}
\end{figure}

Here we consider an interlayer \textbf{k}-SOC as well as interlayer hopping between the $d_{xy}$ and either of the $d_{xz}/d_{yz}$ orbitals. The \textbf{k}-SOC is the spin-dependent part of this process that occurs between states with the same spin in adjacent layers, via the apical oxygen sites as shown in Fig.~\hyperref[interlayer]{4(a)}, which displays the hopping process for the $d_{xz}/d_{xy}$ case.
A schematic illustration of the relevant hopping amplitudes is represented in Fig.~\hyperref[interlayer]{4(b)}, where $t_{pd}^{xy,x}$($t_{pd}^{xy,y}$) is the hopping amplitude between the $d_{xy}$ and $p_{x}$($p_{y}$) orbitals in different layers with equal magnitudes, $t_{pd}^{xy}$. The hopping amplitude occurring purely in the $\hat{z}$ direction between the $d_{xz}$ orbital and the $p_{x}$ at the apical oxygen site is $t_{pd}^{xz,x}$.

We have for the matrix element represented by Fig.~\ref{interlayer},
\begin{equation} \label{amplitude2}
\begin{aligned}
    &\langle{xy,\sigma,\textbf{R}'}|H_{SOC}^{E_{g}} + H_{layer}^{E_{g}}|xz,\sigma,\textbf{R}\rangle = \\[6pt]  &\frac{t_{pd}^{xz,x}}{2(E_{pd} + \frac{\lambda_{p}}{2})(E_{pd} - \lambda_{p})}\biggl[{t_{pd}^{xy,x}}(2E_{pd}-\lambda_{p}) - i\eta_{\sigma}\lambda_{p}{t_{pd}^{xy,y}}\biggr],
\end{aligned}
\end{equation}
from which we see that the real part gives a spin-independent hopping between the orbitals and the imaginary part gives a spin-dependent hopping. Both real and imaginary parts are odd in $z$ since $t_{pd}^{xz,x}$ is odd with respect to reflection about the $xy$ plane. Furthermore, since the real and imaginary parts are proportional to $t_{pd}^{xy,x}$ and $t_{pd}^{xy,y}$ respectively, the real part will be even (odd) in $x (y)$ with the opposite signs for the imaginary part. A $C_{4}$ rotation yields the equivalent result involving $d_{yz}$, but with an opposite even/odd sign dependence with respect to the $x,y$ directions. The result is an effective interlayer hopping,
\begin{equation}
\begin{aligned}
    &H_{layer}^{E_{g}} = -8t_{z}\sum_{\textbf{k}\sigma}\cos{\frac{k_x}{2}}\sin{\frac{k_y}{2}}\sin{\frac{k_z}{2}}c^{\dagger}_{xy,\textbf{k}\sigma}c_{xz,\textbf{k}\sigma} \\ &-8t_{z}\sum_{\textbf{k}\sigma}\sin{\frac{k_x}{2}}\cos{\frac{k_y}{2}}\sin{\frac{k_z}{2}}c^{\dagger}_{xy,\textbf{k}\sigma}c_{yz,\textbf{k}\sigma} + \text{H.c.},
\end{aligned}
\end{equation}
as well as the effective SOC,
\begin{equation}    \label{EgEqn_1}
\begin{aligned}
    H_{SOC}^{E_{g}} =  
    & -8i\lambda^{E_{g}}\sum_{\textbf{k}\sigma\sigma'}\sigma^{z}_{\sigma\sigma'}\sin{\frac{k_x}{2}}\cos{\frac{k_y}{2}}\sin{\frac{k_z}{2}}c^{\dagger}_{xy,\textbf{k}\sigma}c_{xz,\textbf{k}\sigma'}\\
    +&8i\lambda^{E_{g}}\sum_{\textbf{k}\sigma\sigma'}\sigma^{z}_{\sigma\sigma'}\cos{\frac{k_x}{2}}\sin{\frac{k_y}{2}}\sin{\frac{k_z}{2}}c^{\dagger}_{xy,\textbf{k}\sigma}c_{yz,\textbf{k}\sigma'} \\&+ \text{H.c.},
\end{aligned}
\end{equation}
where the effective hopping amplitudes are
\begin{equation}
\begin{aligned}
    t_{z} &= \frac{t_{pd}^{xy}t_{pd}^{xz,x}}{2(E_{pd} + \frac{\lambda_{p}}{2})(E_{pd} - \lambda_{p})}(2E_{pd} - \lambda_{p}), \\[6pt]
    \lambda^{E_{g}} &= \frac{-t_{pd}^{xy}t_{pd}^{xz,x}}{2(E_{pd} + \frac{\lambda_{p}}{2})(E_{pd} - \lambda_{p})}\lambda_{p}.
\end{aligned}
\end{equation}

In summary, we have shown how three different \textbf{k}-SOC terms with distinct symmetries can be generated within a model consisting of the $t_{2g}$ orbitals and oxygen $p$-orbitals with associated on-site SOC. A fit to the density-functional theory (DFT) results of Ref.~[\onlinecite{Veenstra2014PRL}] was carried out within a TB model including \textbf{k}-SOC in Ref.~[\onlinecite{Suh2019}], for which the size of the various \textbf{k}-SOC parameters were all determined to be $\mathcal{O}(1\text{meV})$. However, as discussed in Ref.~[\onlinecite{Suh2019}], through comparison to angle-resolved photoemission spectroscopy (ARPES) measurements \cite{Tamai2019PRX}, it can be seen that the DFT parameters do not accurately account for the size of the SOC, which is enhanced through correlation effects \cite{Tamai2019PRX}. Thus, as mentioned previously, quantifying the values of \textbf{k}-SOC in Sr$_{2}$RuO$_{4}$ requires future studies.

With a microscopic understanding of the origin of these three forms of \textbf{k}-SOC, we next turn to incorporating them into a three-orbital model and study the pairing instabilities that arise in Sr$_{2}$RuO$_{4}$ when the on-site interactions are included.

\section{\label{Five} Application to \protect\Sr}

We apply the shadowed triplet pairing scenario to the unconventional superconductor Sr$_{2}$RuO$_{4}$ \cite{Maeno1994Nature, Mackenzie2003RMP, Kallin2012rpp, Mackenzie2017NPJ} by performing numerical calculations within MF theory for three $t_{2g}$ orbitals. This includes 18 spin-triplet order parameters, described in Eq.~(\ref{OPs}), which are solved self-consistently at zero temperature. Let us take a Hamiltonian, $H = H_{0} + H_{SOC} + H_{pair}$, where the kinetic term, $H_{0}$, is given by
\begin{equation}
H_{0} = \sum_{\textbf{k},\sigma,a} \epsilon^{a}_{\textbf{k}}c_{a,\textbf{k}\sigma}^{\dagger}c_{a,\textbf{k}\sigma} + \sum_{\textbf{k}\sigma}t_{\textbf{k}}c_{yz,\textbf{k}\sigma}^{\dagger}c_{xz,\textbf{k}\sigma} + \text{H.c.},
\end{equation}
with $a$ the orbital index and orbital dispersions, $\displaystyle \epsilon_{\textbf{k}}^{yz/xz} = -2t_{1}\cos{k_{y/x}} - 2t_{2}\cos{k_{x/y}} - \mu_{1d}$, $\epsilon_{\textbf{k}}^{xy} = -2t_{3}(\cos{k_x} + \cos{k_y}) - 4t_{4}\cos{k_x}\cos{k_y} - \mu_{xy}$ and $t_{\textbf{k}} = -4t_{ab}\sin{k_x}\sin{k_y} + 4t_{ab}'\sin{\frac{k_x}{2}}\sin{\frac{k_{y}}{2}}\cos{\frac{k_{z}}{2}}$. The TB parameters, introduced in Ref.~[\onlinecite{lindquist2019distinct}], are ($t_{1}$, $t_{2}$, $t_{3}$, $t_{4}$, $t_{ab}$, $t_{ab}'$, $\mu_{1d}$, $\mu_{xy}$) = (0.45, 0.05, 0.5, 0.2, 0.0025, 0.025, 0.54, 0.64), where all parameters are in units of 2$t_{3}$. The SOC Hamiltonian is
\begin{equation}
    H_{SOC} = H_{SOC}^{A_{1g}} + H_{SOC}^{B_{2g}} + H_{SOC}^{E_{g}},
\end{equation}
where the atomic SOC in the basis of $t_{2g}$ orbitals is $H_{SOC}^{A_{1g}} = i\lambda\sum_{\textbf{k},abc,\sigma\sigma'}\varepsilon_{abc}c^{\dagger}_{a,\textbf{k}\sigma}c_{b,\textbf{k}\sigma'}\sigma_{\sigma\sigma'}^{c}$, and $\varepsilon_{abc}$ is the completely anti-symmetric tensor with $a,b,c = (1,2,3) = (yz,xz,xy)$ representing the $t_{2g}$ orbitals. The resulting TB model is capable of reproducing the experimental FS of Sr$_{2}$RuO$_{4}$ \cite{Mackenzie1996PRL, Bergemann2000PRL, Damascelli2000PRL, Tamai2019PRX} and the pairing term, $H_{pair}$, consists of the attractive channel expressed in terms of the interorbital spin-triplet order parameters in Eq.~(\ref{interactions}).

\begin{figure}[htb!]
\includegraphics[width=86mm]{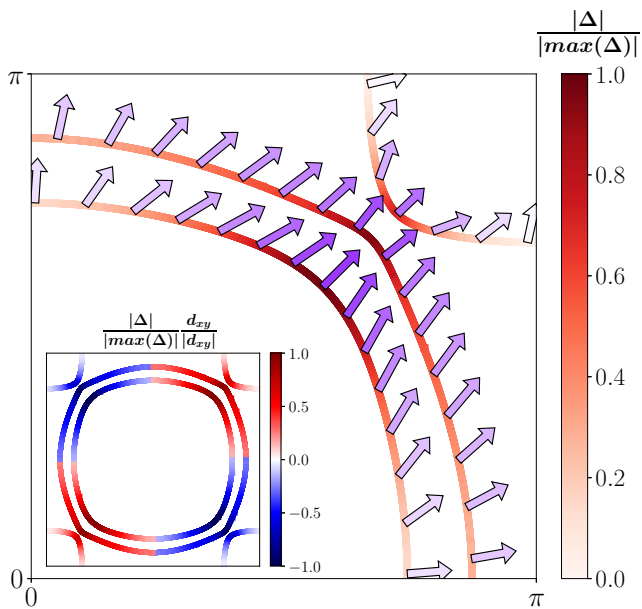}% 
\caption{\label{sidxy}Gap at the FS at $k_{z} = 0$ for a representative $s+id_{xy}$ solution with $\lambda=0.05$, $\lambda^{E_{g}} = 0.005$ and $\lambda^{B_{2g}} = -0.0305$, for which the maximum gap over the FS is $2.4\times10^{-5}$ and the minimum gap is $1.1\times10^{-6}$, where all energies are in units of 2$t_{3}$ (see main text for details). The direction of the arrows indicates the in-plane component of the interorbital spin-triplet $d$-vector associated with the shadowed triplet state, which transforms to intraband pseudospin-singlet pairing on the FS. The length of the arrow indicates the magnitude of the in-plane component; the shorter the arrow, the bigger the $c$-axis component. The inset displays the magnitude of the gap over the FS throughout the full BZ, with the sign of the $d_{xy}$ component of the gap function shown.}
\end{figure}

The MF results are obtained with $J_{H}-U' = 0.7$ and the atomic SOC fixed to $\lambda = 0.05$. With $\lambda^{B_{2g}} = 0$, we obtain a purely $s$-wave solution with $d_{xz/xy}^{x} = d_{yz/xy}^{y} > d_{yz/xz}^{z}$, as shown in Ref.~[\onlinecite{Puetter2012EPL}]. As $\lambda^{B_{2g}}$ reaches an appreciable percentage of $\lambda$, the $d_{xy}$ component becomes non-zero, with orbital MFs $d_{xz/xy}^{y} = d_{yz/xy}^{x}$. Also, there is a relative phase of $\frac{\pi}{2}$ compared to the $s$-component of pairing. This solution is of the form $s + id_{xy}$ in the band basis and has underlying triplet character with a predominantly in-plane $d$-vector involving pairing mostly between the $d_{xz}$ and $d_{xy}$ as well as $d_{yz}$ and $d_{xy}$ orbitals. For $\lambda^{B_{2g}} \approx -0.0305$, the $d_{xy}$ and $s$ components are approximately equal in magnitude. The QP gap is found to be maximum along the $k_{x}=k_{y}$ line, with $\Delta_{max} = 2.4 \times 10^{-5}$ and $\Delta_{min} = 1.1 \times 10^{-6}$. The $s+id_{xy}$ pairing state exists for a range of $\lambda$ and $\lambda^{B_{2g}}$ values around this. Increasing $\lambda^{B_{2g}}$ will eventually lead to a state with only the $d_{xy}$ component. We note that while we have derived $H_{SOC}^{B_{2g}}$ through a perturbative process and $\lambda^{B_{2g}}$ is therefore expected to be smaller than the SOC of oxygen, $\lambda_{p}\approx$ 20meV \cite{oguchi1995PRB}, there should be additional enhancement of the B$_{2g}$ SOC through the coupling to the atomic SOC and orbital hybridization, as well as correlations, which we discussed earlier. We therefore treat $\lambda^{B_{2g}}$ as an effective parameter taking these effects into account, while yielding a FS in agreement with ARPES.

We show the gap on the FS in Fig. \ref{sidxy}, along with arrows indicating the nature of the shadowed triplet at select \textbf{k} points away from the Fermi energy. The SC gap is smallest on the $\alpha$ band along the BZ boundaries, on the order of $1\%$ of the maximum gap or $\mathcal{O}(1\mu$eV), using $2t_{3} \approx 700-800$meV \cite{Kim2018PRL,Roising2019prr}, and can be even smaller as the MF gap is generally overestimated. Furthermore, these deep minima in the gap are robust to changes in the SOC parameters within the region where the $s + id_{xy}$ solution is stabilized. The direction of the arrows indicates the in-plane direction of the $d$-vector, with the length indicating the magnitude of the in-plane component; the shorter the arrow, the bigger the $c$-axis component. Due to the $d_{xy}$ component vanishing along the $a$- and $b$- axis, the pairing solution is composed of $d^{x}_{xz/xy}$ along the $a$-axis and $d^{y}_{yz/xy}$ along the $b$-axis, leading to a $d$-vector that is parallel to the respective axis. The vanishing of the $d_{xy}$ component of the pairing along the $k_{x}/k_{y}$ directions is illustrated in the inset of Fig. \ref{sidxy}.

Within a microscopic theory including Kanamori interactions, the atomic SOC and the B$_{2g}$ and E$_{g}$ \textbf{k}-SOCs derived in Sec.~\hyperref[Four]{IV}, it is possible to stabilize both order parameters of the $s + id_{xy}$ or $d_{xz} + id_{yz}$ type at the FS, depending on the relative size of the E$_{g}$, B$_{2g}$ and atomic SOC strengths. As discussed in Sec.~\hyperref[Three]{III}, these order parameters will appear as intraband pseudospin-singlets on the FS, but with underlying interorbital triplet character originating from the orbital order parameters, $\boldsymbol{d}_{a/b}$. The atomic SOC will stabilize MFs of the form $d^{x}_{xz/xy} = d^{y}_{yz/xy} > d^{z}_{yz/xz}$ \cite{Puetter2012EPL, lindquist2019distinct,Cheung2019PRB,Suh2019}. The B$_{2g}$ SOC, which is given by Eq.~(\ref{B2gEqn_1}), will favor an order parameter that appears with $d_{xy}$ symmetry and underlying $d^{x}_{yz/xy}$ and $d^{y}_{xz/xy}$ triplet character. The E$_{g}$ SOC, given by Eq.~(\ref{EgEqn_1}), will favor a multicomponent order parameter that appears as $d_{xz} + id_{yz}$ \cite{Cheung2019PRB,Suh2019} and have underlying $d^{z}_{xz/xy}$ and $d^{z}_{yz/xy}$ character. By including both the atomic and B$_{2g}$ SOCs, a multicomponent order parameter with $s + id_{xy}$ symmetry can be stabilized, where the relative size of the $s$ and $d_{xy}$ components is determined by the relative sizes of $\lambda$ and $\lambda^{B_{2g}}$. By comparing the ground-state energies, we find that the $d_{xz} + id_{yz}$ solution becomes favorable over the purely $s$-wave solution when $\lambda^{E_{g}}\approx0.015$ and $\lambda^{B_{2g}} = 0$. However, including the B$_{2g}$ SOC by fixing $\lambda^{B_{2g}} = -0.0305$, the critical value at which the $d_{xz} + id_{yz}$ state is stabilized remains approximately the same. Given that the only contribution to the E$_{g}$ SOC involves hopping between different layers, it is reasonable that $\lambda^{B_{2g}}$ is larger than $\lambda^{E_{g}}$, leading to a dominant $s + id_{xy}$ pairing state over $d_{xz} + id_{yz}$.

There has been a growing body of evidence suggesting that any viable order parameter must be a time-reversal symmetry breaking (TRSB) multicomponent order parameter \cite{Luke1998Nature, Xia2006prl, Kidwingira2006science,Grinenko2020}, with an appropriate symmetry that will lead to a jump in the $c_{66}$ elastic modulus at $T_{c}$ \cite{Ghosh2020, Benhabib2020}, and lead to a substantial reduction of the nuclear magnetic resonance (NMR) Knight shift for an in-plane field \cite{Pustogow2019Nature, Ishida2019}. Two even parity proposals are the multicomponent $\{d_{xz},d_{yz}\}$ order parameter in the two-dimensional E$_{g}$ representation \cite{Zutic2005PRL, Suh2019} and the $d_{x^2-y^2} + ig_{xy(x^2-y^2)}$ order parameter which relies on an accidental degeneracy with components from both the B$_{1g}$ and A$_{2g}$ representations \cite{Kivelson2020npj}. The $s + id_{xy}$ state proposed here is a combination of the A$_{1g}$ and B$_{2g}$ representations and generates a sudden change in the shear $c_{66}$ elastic modulus but not $(c_{11}-c_{12})/2$, consistent with the ultrasound data. This is in contrast to an order parameter in the E$_{g}$ representation, leading to a jump also in the $(c_{11}-c_{12})/2$ elastic modulus, which has not been observed experimentally \cite{Ghosh2020,Benhabib2020}. Since other pairing solutions such as $d_{xz/xy}^{y} = -d_{yz/xy}^{x}$ are also found as local minima in MF theory, further experiments and theoretical studies to pin down different order parameters are left for the future.

\section{\label{Six} Summary and Discussion}

In summary, we have studied the microscopic mechanisms of \textbf{k}-SOC and its importance for even-parity spin-triplet pairing in Hund's metals. By taking a simple two-orbital model, we show how a purely interorbital $s$-wave triplet pairing in the orbital basis becomes an intraband pseudospin-singlet pairing with nontrivial momentum dependence near the FS, as well as pseudospin-singlet and -triplet interband pairings which also contain momentum dependence. In the process, we have illustrated the effects of orbital hybridization and SOC on the interorbital spin-triplet pairing state. Applying the idea to Sr$_{2}$RuO$_{4}$, we have derived several forms of $d$-wave \textbf{k}-SOC in the B$_{1g}$, B$_{2g}$ and E$_{g}$ representations, by including the SOC of the oxygen sites within a model consisting of the $t_{2g}$ orbitals and oxygen $2p$ orbitals. While determining the precise size of the various \textbf{k}-SOC parameters in Sr$_{2}$RuO$_{4}$ will require future study, the perturbative approach taken here is an important step in understanding the microscopic origins of these terms in Sr$_{2}$RuO$_{4}$ and other materials. For instance, based on this analysis it is reasonable to expect that the dominant form of \textbf{k}-SOC will likely be from the next-nearest-neighbor in-plane B$_{2g}$ SOC, which will generate a $d_{xy}$ pairing component in addition to the $s$-wave pairing stabilized by the atomic SOC. Subsequently, we have demonstrated the viability of the $s + id_{xy}$ multicomponent solution by including the atomic SOC, B$_{2g}$ and E$_{g}$ SOCs with $t_{2g}$ orbitals. For a range of the three SOC parameters, the existence of the $s + id_{xy}$ state with a predominantly in-plane $d$-vector is generic and independent of details, while competition with other shadowed triplet pairing symmetries depends on the microscopic parameters. The concept we have presented is also applicable to other correlated Hund's metals with significant SOC.

While the pairing solution we have found manifests as a pseudospin-singlet on the FS, the shadowed triplet nature with a predominantly in-plane $d$-vector will be apparent in the presence of finite fields \cite{Yu2018PRB}. As discussed previously in Ref.~[\onlinecite{lindquist2019distinct}], these properties can be confirmed by NMR under uniaxial strain, with an in-plane field applied along both the direction of the strain and perpendicular to it. The $s + id_{xy}$ state we have presented here has essentially the same property that near the $x/y$ directions, where there is mostly $d_{xz}/d_{yz}$ and $d_{xy}$ orbital character, the $d$-vector is parallel to the crystal axes due to the $s$-wave component of pairing and the fact that the $d_{xy}$ component vanishes along those directions. Therefore, under uniaxial strain, there should be a rotation of the average $d$-vector, leading to an anisotropic Knight shift between the $x$ and $y$ directions when the field is a significant fraction of the gap size. Interestingly, the low field behavior is governed by the pairing near the FS, leading to a more isotropic response consistent with the Knight shift of NMR \cite{lindquist2019distinct}.

Going beyond the purely $s$-wave case, an $s + id_{xy}$ pairing naturally explains experiments suggesting a multicomponent order parameter with TRSB and the observed jump in the $c_{66}$ elastic modulus but not the $(c_{11}-c_{12})/2$ modulus \cite{Luke1998Nature, Xia2006prl, Kidwingira2006science,Grinenko2020, Ghosh2020, Benhabib2020}. While $d_{xz}+id_{yz}$ matches the jump in $c_{66}$, the lack of an observed jump in $(c_{11}-c_{12})/2$ is in favor of the $s + id_{xy}$ pairing state over the $d_{xz} + id_{yz}$ state. These two pairing states can also be distinguished due to their different triplet character. The leading order parameter for the $d_{xz} + id_{yz}$ solution corresponds to an out-of-plane $d$-vector, which would yield no rotation under uniaxial strain in contrast to the behavior of the $s + id_{xy}$ state, as discussed above. Such an experiment could provide a test of the $s + id_{xy}$ order parameter arising from the combination of local interactions and \textbf{k}-SOC in light of the lack of many other viable multicomponent alternatives.

In general, two order parameters from different representations with broken time-reversal symmetry implies the presence of two transition temperatures. However, currently there are conflicting experimental data under uniaxial strain regarding this issue: specific heat measurements show no signs of a second transition \cite{Li2019}, while muon spin relaxation measurements do indicate a splitting of $T_{c}$ and the onset of TRSB \cite{Grinenko2020}. Therefore, future experiments under strain will be an important test for a TRSB pairing state. Beyond the experiments under uniaxial strain, it will be important going forward for future experiments to clarify whether the putative gap nodes \cite{NishiZaki2000JPS,Bonalde2000PRL,Lupien2001PRL,Firmo2013PRB,Hassinger2017PRX,sharma2020PNAS} are indeed nodes or deep gap minima, which arise naturally in the pairing solution considered here, as well as the precise location in \textbf{k}-space of these nodes.

\begin{acknowledgments}
We thank Y.-B. Kim for useful discussions. This work was supported by the Natural Sciences and Engineering Research Council of Canada Discovery Grant No. 06089-2016, the Center for Quantum Materials at
the University of Toronto, the Canadian Institute for
Advanced Research, and the Canada Research Chairs Program. Computations were performed on the
Niagara supercomputer at the SciNet HPC Consortium.
SciNet is funded by: the Canada Foundation for Innovation
under the auspices of Compute Canada; the Government
of Ontario; Ontario Research Fund - Research Excellence;
and the University of Toronto.
\end{acknowledgments}

\bibliography{main}% Produces the bibliography via BibTeX.

\begin{thebibliography}{55}%
\makeatletter
\providecommand \@ifxundefined [1]{%
 \@ifx{#1\undefined}
}%
\providecommand \@ifnum [1]{%
 \ifnum #1\expandafter \@firstoftwo
 \else \expandafter \@secondoftwo
 \fi
}%
\providecommand \@ifx [1]{%
 \ifx #1\expandafter \@firstoftwo
 \else \expandafter \@secondoftwo
 \fi
}%
\providecommand \natexlab [1]{#1}%
\providecommand \enquote  [1]{``#1''}%
\providecommand \bibnamefont  [1]{#1}%
\providecommand \bibfnamefont [1]{#1}%
\providecommand \citenamefont [1]{#1}%
\providecommand \href@noop [0]{\@secondoftwo}%
\providecommand \href [0]{\begingroup \@sanitize@url \@href}%
\providecommand \@href[1]{\@@startlink{#1}\@@href}%
\providecommand \@@href[1]{\endgroup#1\@@endlink}%
\providecommand \@sanitize@url [0]{\catcode `\\12\catcode `\$12\catcode
  `\&12\catcode `\#12\catcode `\^12\catcode `\_12\catcode `\%12\relax}%
\providecommand \@@startlink[1]{}%
\providecommand \@@endlink[0]{}%
\providecommand \url  [0]{\begingroup\@sanitize@url \@url }%
\providecommand \@url [1]{\endgroup\@href {#1}{\urlprefix }}%
\providecommand \urlprefix  [0]{URL }%
\providecommand \Eprint [0]{\href }%
\providecommand \doibase [0]{https://doi.org/}%
\providecommand \selectlanguage [0]{\@gobble}%
\providecommand \bibinfo  [0]{\@secondoftwo}%
\providecommand \bibfield  [0]{\@secondoftwo}%
\providecommand \translation [1]{[#1]}%
\providecommand \BibitemOpen [0]{}%
\providecommand \bibitemStop [0]{}%
\providecommand \bibitemNoStop [0]{.\EOS\space}%
\providecommand \EOS [0]{\spacefactor3000\relax}%
\providecommand \BibitemShut  [1]{\csname bibitem#1\endcsname}%
\let\auto@bib@innerbib\@empty
%</preamble>
\bibitem [{\citenamefont {Klejnberg}\ and\ \citenamefont
  {Spalek}(1999)}]{klejnberg1999hund}%
  \BibitemOpen
  \bibfield  {author} {\bibinfo {author} {\bibfnamefont {A.}~\bibnamefont
  {Klejnberg}}\ and\ \bibinfo {author} {\bibfnamefont {J.}~\bibnamefont
  {Spalek}},\ }\bibfield  {title} {\bibinfo {title} {Hund's rule coupling as
  the microscopic origin of the spin-triplet pairing in a correlated and
  degenerate band system},\ }\href
  {https://doi.org/10.1088/0953-8984/11/34/307} {\bibfield  {journal} {\bibinfo
   {journal} {J. Phys.: Condens. Matter}\ }\textbf {\bibinfo {volume} {11}},\
  \bibinfo {pages} {6553} (\bibinfo {year} {1999})}\BibitemShut {NoStop}%
\bibitem [{\citenamefont {Dai}\ \emph {et~al.}(2008)\citenamefont {Dai},
  \citenamefont {Fang}, \citenamefont {Zhou},\ and\ \citenamefont
  {Zhang}}]{Dai2008PRL}%
  \BibitemOpen
  \bibfield  {author} {\bibinfo {author} {\bibfnamefont {X.}~\bibnamefont
  {Dai}}, \bibinfo {author} {\bibfnamefont {Z.}~\bibnamefont {Fang}}, \bibinfo
  {author} {\bibfnamefont {Y.}~\bibnamefont {Zhou}},\ and\ \bibinfo {author}
  {\bibfnamefont {F.-C.}\ \bibnamefont {Zhang}},\ }\bibfield  {title} {\bibinfo
  {title} {Even {P}arity, {O}rbital {S}inglet, and {S}pin {T}riplet {P}airing
  for {S}uperconducting
  {${\mathrm{LaFeAsO}}_{1\ensuremath{-}x}{\mathrm{F}}_{x}$}},\ }\href
  {https://doi.org/10.1103/PhysRevLett.101.057008} {\bibfield  {journal}
  {\bibinfo  {journal} {Phys. Rev. Lett.}\ }\textbf {\bibinfo {volume} {101}},\
  \bibinfo {pages} {057008} (\bibinfo {year} {2008})}\BibitemShut {NoStop}%
\bibitem [{\citenamefont {Puetter}\ and\ \citenamefont
  {Kee}(2012)}]{Puetter2012EPL}%
  \BibitemOpen
  \bibfield  {author} {\bibinfo {author} {\bibfnamefont {C.~M.}\ \bibnamefont
  {Puetter}}\ and\ \bibinfo {author} {\bibfnamefont {H.-Y.}\ \bibnamefont
  {Kee}},\ }\bibfield  {title} {\bibinfo {title} {Identifying spin-triplet
  pairing in spin-orbit coupled multi-band superconductors},\ }\href
  {https://doi.org/10.1209/0295-5075/98/27010} {\bibfield  {journal} {\bibinfo
  {journal} {EPL (Europhysics Letters)}\ }\textbf {\bibinfo {volume} {98}},\
  \bibinfo {pages} {27010} (\bibinfo {year} {2012})}\BibitemShut {NoStop}%
\bibitem [{\citenamefont {Hoshino}\ and\ \citenamefont
  {Werner}(2015)}]{Hoshino2015PRL}%
  \BibitemOpen
  \bibfield  {author} {\bibinfo {author} {\bibfnamefont {S.}~\bibnamefont
  {Hoshino}}\ and\ \bibinfo {author} {\bibfnamefont {P.}~\bibnamefont
  {Werner}},\ }\bibfield  {title} {\bibinfo {title} {Superconductivity from
  {E}merging {M}agnetic {M}oments},\ }\href
  {https://doi.org/10.1103/PhysRevLett.115.247001} {\bibfield  {journal}
  {\bibinfo  {journal} {Phys. Rev. Lett.}\ }\textbf {\bibinfo {volume} {115}},\
  \bibinfo {pages} {247001} (\bibinfo {year} {2015})}\BibitemShut {NoStop}%
\bibitem [{\citenamefont {Hoshino}\ and\ \citenamefont
  {Werner}(2016)}]{Hoshino2016PRB}%
  \BibitemOpen
  \bibfield  {author} {\bibinfo {author} {\bibfnamefont {S.}~\bibnamefont
  {Hoshino}}\ and\ \bibinfo {author} {\bibfnamefont {P.}~\bibnamefont
  {Werner}},\ }\bibfield  {title} {\bibinfo {title} {Electronic orders in
  multiorbital {H}ubbard models with lifted orbital degeneracy},\ }\href
  {https://doi.org/10.1103/PhysRevB.93.155161} {\bibfield  {journal} {\bibinfo
  {journal} {Phys. Rev. B}\ }\textbf {\bibinfo {volume} {93}},\ \bibinfo
  {pages} {155161} (\bibinfo {year} {2016})}\BibitemShut {NoStop}%
\bibitem [{\citenamefont {Gingras}\ \emph {et~al.}(2019)\citenamefont
  {Gingras}, \citenamefont {Nourafkan}, \citenamefont {Tremblay},\ and\
  \citenamefont {C\^ot\'e}}]{Gingras2019PRL}%
  \BibitemOpen
  \bibfield  {author} {\bibinfo {author} {\bibfnamefont {O.}~\bibnamefont
  {Gingras}}, \bibinfo {author} {\bibfnamefont {R.}~\bibnamefont {Nourafkan}},
  \bibinfo {author} {\bibfnamefont {A.-M.~S.}\ \bibnamefont {Tremblay}},\ and\
  \bibinfo {author} {\bibfnamefont {M.}~\bibnamefont {C\^ot\'e}},\ }\bibfield
  {title} {\bibinfo {title} {Superconducting {S}ymmetries of
  {${\mathrm{Sr}}_{2}{\mathrm{RuO}}_{4}$} from {F}irst-{P}rinciples
  {E}lectronic {S}tructure},\ }\href
  {https://doi.org/10.1103/PhysRevLett.123.217005} {\bibfield  {journal}
  {\bibinfo  {journal} {Phys. Rev. Lett.}\ }\textbf {\bibinfo {volume} {123}},\
  \bibinfo {pages} {217005} (\bibinfo {year} {2019})}\BibitemShut {NoStop}%
\bibitem [{\citenamefont {Vafek}\ and\ \citenamefont
  {Chubukov}(2017)}]{vafek2017hund}%
  \BibitemOpen
  \bibfield  {author} {\bibinfo {author} {\bibfnamefont {O.}~\bibnamefont
  {Vafek}}\ and\ \bibinfo {author} {\bibfnamefont {A.~V.}\ \bibnamefont
  {Chubukov}},\ }\bibfield  {title} {\bibinfo {title} {Hund {I}nteraction,
  {S}pin-{O}rbit {C}oupling, and the {M}echanism of {S}uperconductivity in
  {S}trongly {H}ole-{D}oped {I}ron {P}nictides},\ }\href
  {https://link.aps.org/doi/10.1103/PhysRevLett.118.087003} {\bibfield
  {journal} {\bibinfo  {journal} {Phys. Rev. Lett.}\ }\textbf {\bibinfo
  {volume} {118}},\ \bibinfo {pages} {087003} (\bibinfo {year}
  {2017})}\BibitemShut {NoStop}%
\bibitem [{\citenamefont {Joynt}\ and\ \citenamefont
  {Taillefer}(2002)}]{Joynt2002RMP}%
  \BibitemOpen
  \bibfield  {author} {\bibinfo {author} {\bibfnamefont {R.}~\bibnamefont
  {Joynt}}\ and\ \bibinfo {author} {\bibfnamefont {L.}~\bibnamefont
  {Taillefer}},\ }\bibfield  {title} {\bibinfo {title} {The superconducting
  phases of {${\mathrm{UPt}}_{3}$}},\ }\href
  {https://doi.org/10.1103/RevModPhys.74.235} {\bibfield  {journal} {\bibinfo
  {journal} {Rev. Mod. Phys.}\ }\textbf {\bibinfo {volume} {74}},\ \bibinfo
  {pages} {235} (\bibinfo {year} {2002})}\BibitemShut {NoStop}%
\bibitem [{\citenamefont {Kallin}\ and\ \citenamefont
  {Berlinsky}(2016)}]{Kallin2016}%
  \BibitemOpen
  \bibfield  {author} {\bibinfo {author} {\bibfnamefont {C.}~\bibnamefont
  {Kallin}}\ and\ \bibinfo {author} {\bibfnamefont {J.}~\bibnamefont
  {Berlinsky}},\ }\bibfield  {title} {\bibinfo {title} {Chiral
  superconductors},\ }\href {https://doi.org/10.1088/0034-4885/79/5/054502}
  {\bibfield  {journal} {\bibinfo  {journal} {Rep. Prog. Phys.}\ }\textbf
  {\bibinfo {volume} {79}},\ \bibinfo {pages} {054502} (\bibinfo {year}
  {2016})}\BibitemShut {NoStop}%
\bibitem [{\citenamefont {Fu}\ and\ \citenamefont {Kane}(2008)}]{Fu2008PRL}%
  \BibitemOpen
  \bibfield  {author} {\bibinfo {author} {\bibfnamefont {L.}~\bibnamefont
  {Fu}}\ and\ \bibinfo {author} {\bibfnamefont {C.~L.}\ \bibnamefont {Kane}},\
  }\bibfield  {title} {\bibinfo {title} {Superconducting {P}roximity {E}ffect
  and {M}ajorana {F}ermions at the {S}urface of a {T}opological {I}nsulator},\
  }\href {https://doi.org/10.1103/PhysRevLett.100.096407} {\bibfield  {journal}
  {\bibinfo  {journal} {Phys. Rev. Lett.}\ }\textbf {\bibinfo {volume} {100}},\
  \bibinfo {pages} {096407} (\bibinfo {year} {2008})}\BibitemShut {NoStop}%
\bibitem [{\citenamefont {Fujimoto}(2008)}]{Fujimoto2008PRB}%
  \BibitemOpen
  \bibfield  {author} {\bibinfo {author} {\bibfnamefont {S.}~\bibnamefont
  {Fujimoto}},\ }\bibfield  {title} {\bibinfo {title} {Topological order and
  non-{A}belian statistics in noncentrosymmetric $s$-wave superconductors},\
  }\href {https://doi.org/10.1103/PhysRevB.77.220501} {\bibfield  {journal}
  {\bibinfo  {journal} {Phys. Rev. B}\ }\textbf {\bibinfo {volume} {77}},\
  \bibinfo {pages} {220501(R)} (\bibinfo {year} {2008})}\BibitemShut {NoStop}%
\bibitem [{\citenamefont {Fu}\ and\ \citenamefont {Kane}(2009)}]{Fu2009PRB}%
  \BibitemOpen
  \bibfield  {author} {\bibinfo {author} {\bibfnamefont {L.}~\bibnamefont
  {Fu}}\ and\ \bibinfo {author} {\bibfnamefont {C.~L.}\ \bibnamefont {Kane}},\
  }\bibfield  {title} {\bibinfo {title} {Josephson current and noise at a
  superconductor/quantum-spin-{H}all-insulator/superconductor junction},\
  }\href {https://doi.org/10.1103/PhysRevB.79.161408} {\bibfield  {journal}
  {\bibinfo  {journal} {Phys. Rev. B}\ }\textbf {\bibinfo {volume} {79}},\
  \bibinfo {pages} {161408(R)} (\bibinfo {year} {2009})}\BibitemShut {NoStop}%
\bibitem [{\citenamefont {Chen}\ and\ \citenamefont {Kee}(2014)}]{YigePRB2014}%
  \BibitemOpen
  \bibfield  {author} {\bibinfo {author} {\bibfnamefont {Y.}~\bibnamefont
  {Chen}}\ and\ \bibinfo {author} {\bibfnamefont {H.-Y.}\ \bibnamefont {Kee}},\
  }\bibfield  {title} {\bibinfo {title} {Topological phases in iridium oxide
  superlattices: Quantized anomalous charge or valley {H}all insulators},\
  }\href {https://doi.org/10.1103/PhysRevB.90.195145} {\bibfield  {journal}
  {\bibinfo  {journal} {Phys. Rev. B}\ }\textbf {\bibinfo {volume} {90}},\
  \bibinfo {pages} {195145} (\bibinfo {year} {2014})}\BibitemShut {NoStop}%
\bibitem [{\citenamefont {Maeno}\ \emph {et~al.}(1994)\citenamefont {Maeno},
  \citenamefont {Hashimoto}, \citenamefont {Yoshida}, \citenamefont
  {Nishizaki}, \citenamefont {Fujita}, \citenamefont {Bednorz},\ and\
  \citenamefont {Lichtenberg}}]{Maeno1994Nature}%
  \BibitemOpen
  \bibfield  {author} {\bibinfo {author} {\bibfnamefont {Y.}~\bibnamefont
  {Maeno}}, \bibinfo {author} {\bibfnamefont {H.}~\bibnamefont {Hashimoto}},
  \bibinfo {author} {\bibfnamefont {K.}~\bibnamefont {Yoshida}}, \bibinfo
  {author} {\bibfnamefont {S.}~\bibnamefont {Nishizaki}}, \bibinfo {author}
  {\bibfnamefont {T.}~\bibnamefont {Fujita}}, \bibinfo {author} {\bibfnamefont
  {J.~G.}\ \bibnamefont {Bednorz}},\ and\ \bibinfo {author} {\bibfnamefont
  {F.}~\bibnamefont {Lichtenberg}},\ }\bibfield  {title} {\bibinfo {title}
  {Superconductivity in a layered perovskite without copper},\ }\href
  {https://doi.org/10.1038/372532a0} {\bibfield  {journal} {\bibinfo  {journal}
  {Nature (London)}\ }\textbf {\bibinfo {volume} {372}},\ \bibinfo {pages}
  {532} (\bibinfo {year} {1994})}\BibitemShut {NoStop}%
\bibitem [{\citenamefont {Mackenzie}\ and\ \citenamefont
  {Maeno}(2003)}]{Mackenzie2003RMP}%
  \BibitemOpen
  \bibfield  {author} {\bibinfo {author} {\bibfnamefont {A.~P.}\ \bibnamefont
  {Mackenzie}}\ and\ \bibinfo {author} {\bibfnamefont {Y.}~\bibnamefont
  {Maeno}},\ }\bibfield  {title} {\bibinfo {title} {The superconductivity of
  {${\mathrm{Sr}}_{2}{\mathrm{RuO}}_{4}$} and the physics of spin-triplet
  pairing},\ }\href {https://doi.org/10.1103/RevModPhys.75.657} {\bibfield
  {journal} {\bibinfo  {journal} {Rev. Mod. Phys.}\ }\textbf {\bibinfo {volume}
  {75}},\ \bibinfo {pages} {657} (\bibinfo {year} {2003})}\BibitemShut
  {NoStop}%
\bibitem [{\citenamefont {Kallin}(2012)}]{Kallin2012rpp}%
  \BibitemOpen
  \bibfield  {author} {\bibinfo {author} {\bibfnamefont {C.}~\bibnamefont
  {Kallin}},\ }\bibfield  {title} {\bibinfo {title} {{Chiral p-wave order in
  Sr$_2$RuO$_4$}},\ }\href {https://doi.org/10.1088/0034-4885/75/4/042501}
  {\bibfield  {journal} {\bibinfo  {journal} {Rep. Prog. Phys.}\ }\textbf
  {\bibinfo {volume} {75}},\ \bibinfo {pages} {042501} (\bibinfo {year}
  {2012})}\BibitemShut {NoStop}%
\bibitem [{\citenamefont {Mackenzie}\ \emph {et~al.}(2017)\citenamefont
  {Mackenzie}, \citenamefont {Scaffidi}, \citenamefont {Hicks},\ and\
  \citenamefont {Maeno}}]{Mackenzie2017NPJ}%
  \BibitemOpen
  \bibfield  {author} {\bibinfo {author} {\bibfnamefont {A.~P.}\ \bibnamefont
  {Mackenzie}}, \bibinfo {author} {\bibfnamefont {T.}~\bibnamefont {Scaffidi}},
  \bibinfo {author} {\bibfnamefont {C.~W.}\ \bibnamefont {Hicks}},\ and\
  \bibinfo {author} {\bibfnamefont {Y.}~\bibnamefont {Maeno}},\ }\bibfield
  {title} {\bibinfo {title} {Even odder after twenty-three years: the
  superconducting order parameter puzzle of
  {${\mathrm{Sr}}_{2}{\mathrm{RuO}}_{4}$}},\ }\href
  {https://doi.org/10.1038/s41535-017-0045-4} {\bibfield  {journal} {\bibinfo
  {journal} {npj Quantum Mater.}\ }\textbf {\bibinfo {volume} {2}},\ \bibinfo
  {pages} {40} (\bibinfo {year} {2017})}\BibitemShut {NoStop}%
\bibitem [{\citenamefont {Cheung}\ and\ \citenamefont
  {Agterberg}(2019)}]{Cheung2019PRB}%
  \BibitemOpen
  \bibfield  {author} {\bibinfo {author} {\bibfnamefont {A.~K.~C.}\
  \bibnamefont {Cheung}}\ and\ \bibinfo {author} {\bibfnamefont {D.~F.}\
  \bibnamefont {Agterberg}},\ }\bibfield  {title} {\bibinfo {title}
  {Superconductivity in the presence of spin-orbit interactions stabilized by
  {H}und coupling},\ }\href {https://doi.org/10.1103/PhysRevB.99.024516}
  {\bibfield  {journal} {\bibinfo  {journal} {Phys. Rev. B}\ }\textbf {\bibinfo
  {volume} {99}},\ \bibinfo {pages} {024516} (\bibinfo {year}
  {2019})}\BibitemShut {NoStop}%
\bibitem [{\citenamefont {Ramires}\ and\ \citenamefont
  {Sigrist}(2019)}]{Ramires2019PRB}%
  \BibitemOpen
  \bibfield  {author} {\bibinfo {author} {\bibfnamefont {A.}~\bibnamefont
  {Ramires}}\ and\ \bibinfo {author} {\bibfnamefont {M.}~\bibnamefont
  {Sigrist}},\ }\bibfield  {title} {\bibinfo {title} {Superconducting order
  parameter of {${\mathrm{Sr}}_{2}{\mathrm{RuO}}_{4}$}: A microscopic
  perspective},\ }\href {https://doi.org/10.1103/PhysRevB.100.104501}
  {\bibfield  {journal} {\bibinfo  {journal} {Phys. Rev. B}\ }\textbf {\bibinfo
  {volume} {100}},\ \bibinfo {pages} {104501} (\bibinfo {year}
  {2019})}\BibitemShut {NoStop}%
\bibitem [{\citenamefont {Suh}\ \emph {et~al.}(2020)\citenamefont {Suh},
  \citenamefont {Menke}, \citenamefont {Brydon}, \citenamefont {Timm},
  \citenamefont {Ramires},\ and\ \citenamefont {Agterberg}}]{Suh2019}%
  \BibitemOpen
  \bibfield  {author} {\bibinfo {author} {\bibfnamefont {H.~G.}\ \bibnamefont
  {Suh}}, \bibinfo {author} {\bibfnamefont {H.}~\bibnamefont {Menke}}, \bibinfo
  {author} {\bibfnamefont {P.~M.~R.}\ \bibnamefont {Brydon}}, \bibinfo {author}
  {\bibfnamefont {C.}~\bibnamefont {Timm}}, \bibinfo {author} {\bibfnamefont
  {A.}~\bibnamefont {Ramires}},\ and\ \bibinfo {author} {\bibfnamefont {D.~F.}\
  \bibnamefont {Agterberg}},\ }\bibfield  {title} {\bibinfo {title}
  {Stabilizing even-parity chiral superconductivity in
  {${\mathrm{Sr}}_{2}{\mathrm{RuO}}_{4}$}},\ }\href
  {https://link.aps.org/doi/10.1103/PhysRevResearch.2.032023} {\bibfield
  {journal} {\bibinfo  {journal} {Phys. Rev. Res.}\ }\textbf {\bibinfo {volume}
  {2}},\ \bibinfo {pages} {032023} (\bibinfo {year} {2020})}\BibitemShut
  {NoStop}%
\bibitem [{\citenamefont {Lindquist}\ and\ \citenamefont
  {Kee}(2020)}]{lindquist2019distinct}%
  \BibitemOpen
  \bibfield  {author} {\bibinfo {author} {\bibfnamefont {A.~W.}\ \bibnamefont
  {Lindquist}}\ and\ \bibinfo {author} {\bibfnamefont {H.-Y.}\ \bibnamefont
  {Kee}},\ }\bibfield  {title} {\bibinfo {title} {Distinct reduction of
  {K}night shift in superconducting state of
  {${\mathrm{Sr}}_{2}{\mathrm{RuO}}_{4}$} under uniaxial strain},\ }\href
  {https://doi.org/10.1103/PhysRevResearch.2.032055} {\bibfield  {journal}
  {\bibinfo  {journal} {Phys. Rev. Res.}\ }\textbf {\bibinfo {volume} {2}},\
  \bibinfo {pages} {032055} (\bibinfo {year} {2020})}\BibitemShut {NoStop}%
\bibitem [{\citenamefont {Pavarini}\ and\ \citenamefont
  {Mazin}(2006)}]{Pavarini2006PRB}%
  \BibitemOpen
  \bibfield  {author} {\bibinfo {author} {\bibfnamefont {E.}~\bibnamefont
  {Pavarini}}\ and\ \bibinfo {author} {\bibfnamefont {I.~I.}\ \bibnamefont
  {Mazin}},\ }\bibfield  {title} {\bibinfo {title} {First-principles study of
  spin-orbit effects and {NMR} in
  {${\mathrm{Sr}}_{2}\mathrm{Ru}{\mathrm{O}}_{4}$}},\ }\href
  {https://doi.org/10.1103/PhysRevB.74.035115} {\bibfield  {journal} {\bibinfo
  {journal} {Phys. Rev. B}\ }\textbf {\bibinfo {volume} {74}},\ \bibinfo
  {pages} {035115} (\bibinfo {year} {2006})}\BibitemShut {NoStop}%
\bibitem [{\citenamefont {Haverkort}\ \emph {et~al.}(2008)\citenamefont
  {Haverkort}, \citenamefont {Elfimov}, \citenamefont {Tjeng}, \citenamefont
  {Sawatzky},\ and\ \citenamefont {Damascelli}}]{Haverkort2008PRL}%
  \BibitemOpen
  \bibfield  {author} {\bibinfo {author} {\bibfnamefont {M.~W.}\ \bibnamefont
  {Haverkort}}, \bibinfo {author} {\bibfnamefont {I.~S.}\ \bibnamefont
  {Elfimov}}, \bibinfo {author} {\bibfnamefont {L.~H.}\ \bibnamefont {Tjeng}},
  \bibinfo {author} {\bibfnamefont {G.~A.}\ \bibnamefont {Sawatzky}},\ and\
  \bibinfo {author} {\bibfnamefont {A.}~\bibnamefont {Damascelli}},\ }\bibfield
   {title} {\bibinfo {title} {Strong {S}pin-{O}rbit {C}oupling {E}ffects on the
  {F}ermi {S}urface of {${\mathrm{Sr}}_{2}{\mathrm{RuO}}_{4}$} and
  {${\mathrm{Sr}}_{2}{\mathrm{RhO}}_{4}$}},\ }\href
  {https://doi.org/10.1103/PhysRevLett.101.026406} {\bibfield  {journal}
  {\bibinfo  {journal} {Phys. Rev. Lett.}\ }\textbf {\bibinfo {volume} {101}},\
  \bibinfo {pages} {026406} (\bibinfo {year} {2008})}\BibitemShut {NoStop}%
\bibitem [{\citenamefont {Rozbicki}\ \emph {et~al.}(2011)\citenamefont
  {Rozbicki}, \citenamefont {Annett}, \citenamefont {Souquet},\ and\
  \citenamefont {Mackenzie}}]{Rozbicki2011JPCM}%
  \BibitemOpen
  \bibfield  {author} {\bibinfo {author} {\bibfnamefont {E.~J.}\ \bibnamefont
  {Rozbicki}}, \bibinfo {author} {\bibfnamefont {J.~F.}\ \bibnamefont
  {Annett}}, \bibinfo {author} {\bibfnamefont {J.-R.}\ \bibnamefont
  {Souquet}},\ and\ \bibinfo {author} {\bibfnamefont {A.~P.}\ \bibnamefont
  {Mackenzie}},\ }\bibfield  {title} {\bibinfo {title} {Spin{\textendash}orbit
  coupling and k-dependent {Z}eeman splitting in strontium ruthenate},\ }\href
  {https://doi.org/10.1088/0953-8984/23/9/094201} {\bibfield  {journal}
  {\bibinfo  {journal} {J. Phys.: Condens. Matter}\ }\textbf {\bibinfo {volume}
  {23}},\ \bibinfo {pages} {094201} (\bibinfo {year} {2011})}\BibitemShut
  {NoStop}%
\bibitem [{\citenamefont {Veenstra}\ \emph {et~al.}(2014)\citenamefont
  {Veenstra}, \citenamefont {Zhu}, \citenamefont {Raichle}, \citenamefont
  {Ludbrook}, \citenamefont {Nicolaou}, \citenamefont {Slomski}, \citenamefont
  {Landolt}, \citenamefont {Kittaka}, \citenamefont {Maeno}, \citenamefont
  {Dil}, \citenamefont {Elfimov}, \citenamefont {Haverkort},\ and\
  \citenamefont {Damascelli}}]{Veenstra2014PRL}%
  \BibitemOpen
  \bibfield  {author} {\bibinfo {author} {\bibfnamefont {C.~N.}\ \bibnamefont
  {Veenstra}}, \bibinfo {author} {\bibfnamefont {Z.-H.}\ \bibnamefont {Zhu}},
  \bibinfo {author} {\bibfnamefont {M.}~\bibnamefont {Raichle}}, \bibinfo
  {author} {\bibfnamefont {B.~M.}\ \bibnamefont {Ludbrook}}, \bibinfo {author}
  {\bibfnamefont {A.}~\bibnamefont {Nicolaou}}, \bibinfo {author}
  {\bibfnamefont {B.}~\bibnamefont {Slomski}}, \bibinfo {author} {\bibfnamefont
  {G.}~\bibnamefont {Landolt}}, \bibinfo {author} {\bibfnamefont
  {S.}~\bibnamefont {Kittaka}}, \bibinfo {author} {\bibfnamefont
  {Y.}~\bibnamefont {Maeno}}, \bibinfo {author} {\bibfnamefont {J.~H.}\
  \bibnamefont {Dil}}, \bibinfo {author} {\bibfnamefont {I.~S.}\ \bibnamefont
  {Elfimov}}, \bibinfo {author} {\bibfnamefont {M.~W.}\ \bibnamefont
  {Haverkort}},\ and\ \bibinfo {author} {\bibfnamefont {A.}~\bibnamefont
  {Damascelli}},\ }\bibfield  {title} {\bibinfo {title} {Spin-{O}rbital
  {E}ntanglement and the {B}reakdown of {S}inglets and {T}riplets in
  {${\mathrm{Sr}}_{2}{\mathrm{RuO}}_{4}$} {R}evealed by {S}pin- and
  {A}ngle-{R}esolved {P}hotoemission {S}pectroscopy},\ }\href
  {https://doi.org/10.1103/PhysRevLett.112.127002} {\bibfield  {journal}
  {\bibinfo  {journal} {Phys. Rev. Lett.}\ }\textbf {\bibinfo {volume} {112}},\
  \bibinfo {pages} {127002} (\bibinfo {year} {2014})}\BibitemShut {NoStop}%
\bibitem [{\citenamefont {Tamai}\ \emph {et~al.}(2019)\citenamefont {Tamai},
  \citenamefont {Zingl}, \citenamefont {Rozbicki}, \citenamefont {Cappelli},
  \citenamefont {Ricc\`o}, \citenamefont {de~la Torre}, \citenamefont
  {McKeown~Walker}, \citenamefont {Bruno}, \citenamefont {King}, \citenamefont
  {Meevasana}, \citenamefont {Shi}, \citenamefont
  {Radovi\ifmmode~\acute{c}\else \'{c}\fi{}}, \citenamefont {Plumb},
  \citenamefont {Gibbs}, \citenamefont {Mackenzie}, \citenamefont {Berthod},
  \citenamefont {Strand}, \citenamefont {Kim}, \citenamefont {Georges},\ and\
  \citenamefont {Baumberger}}]{Tamai2019PRX}%
  \BibitemOpen
  \bibfield  {author} {\bibinfo {author} {\bibfnamefont {A.}~\bibnamefont
  {Tamai}}, \bibinfo {author} {\bibfnamefont {M.}~\bibnamefont {Zingl}},
  \bibinfo {author} {\bibfnamefont {E.}~\bibnamefont {Rozbicki}}, \bibinfo
  {author} {\bibfnamefont {E.}~\bibnamefont {Cappelli}}, \bibinfo {author}
  {\bibfnamefont {S.}~\bibnamefont {Ricc\`o}}, \bibinfo {author} {\bibfnamefont
  {A.}~\bibnamefont {de~la Torre}}, \bibinfo {author} {\bibfnamefont
  {S.}~\bibnamefont {McKeown~Walker}}, \bibinfo {author} {\bibfnamefont
  {F.~Y.}\ \bibnamefont {Bruno}}, \bibinfo {author} {\bibfnamefont {P.~D.~C.}\
  \bibnamefont {King}}, \bibinfo {author} {\bibfnamefont {W.}~\bibnamefont
  {Meevasana}}, \bibinfo {author} {\bibfnamefont {M.}~\bibnamefont {Shi}},
  \bibinfo {author} {\bibfnamefont {M.}~\bibnamefont
  {Radovi\ifmmode~\acute{c}\else \'{c}\fi{}}}, \bibinfo {author} {\bibfnamefont
  {N.~C.}\ \bibnamefont {Plumb}}, \bibinfo {author} {\bibfnamefont {A.~S.}\
  \bibnamefont {Gibbs}}, \bibinfo {author} {\bibfnamefont {A.~P.}\ \bibnamefont
  {Mackenzie}}, \bibinfo {author} {\bibfnamefont {C.}~\bibnamefont {Berthod}},
  \bibinfo {author} {\bibfnamefont {H.~U.~R.}\ \bibnamefont {Strand}}, \bibinfo
  {author} {\bibfnamefont {M.}~\bibnamefont {Kim}}, \bibinfo {author}
  {\bibfnamefont {A.}~\bibnamefont {Georges}},\ and\ \bibinfo {author}
  {\bibfnamefont {F.}~\bibnamefont {Baumberger}},\ }\bibfield  {title}
  {\bibinfo {title} {High-{R}esolution {P}hotoemission on
  {${\mathrm{Sr}}_{2}{\mathrm{RuO}}_{4}$} {R}eveals {C}orrelation-{E}nhanced
  {E}ffective {S}pin-{O}rbit {C}oupling and {D}ominantly {L}ocal
  {S}elf-{E}nergies},\ }\href {https://doi.org/10.1103/PhysRevX.9.021048}
  {\bibfield  {journal} {\bibinfo  {journal} {Phys. Rev. X}\ }\textbf {\bibinfo
  {volume} {9}},\ \bibinfo {pages} {021048} (\bibinfo {year}
  {2019})}\BibitemShut {NoStop}%
\bibitem [{\citenamefont {Mravlje}\ \emph {et~al.}(2011)\citenamefont
  {Mravlje}, \citenamefont {Aichhorn}, \citenamefont {Miyake}, \citenamefont
  {Haule}, \citenamefont {Kotliar},\ and\ \citenamefont
  {Georges}}]{mravlje2011PRL}%
  \BibitemOpen
  \bibfield  {author} {\bibinfo {author} {\bibfnamefont {J.}~\bibnamefont
  {Mravlje}}, \bibinfo {author} {\bibfnamefont {M.}~\bibnamefont {Aichhorn}},
  \bibinfo {author} {\bibfnamefont {T.}~\bibnamefont {Miyake}}, \bibinfo
  {author} {\bibfnamefont {K.}~\bibnamefont {Haule}}, \bibinfo {author}
  {\bibfnamefont {G.}~\bibnamefont {Kotliar}},\ and\ \bibinfo {author}
  {\bibfnamefont {A.}~\bibnamefont {Georges}},\ }\bibfield  {title} {\bibinfo
  {title} {Coherence-{I}ncoherence {C}rossover and the {M}ass-{R}enormalization
  {P}uzzles in {${\mathrm{Sr}}_{2}{\mathrm{RuO}}_{4}$}},\ }\href
  {https://link.aps.org/doi/10.1103/PhysRevLett.106.096401} {\bibfield
  {journal} {\bibinfo  {journal} {Phys. Rev. Lett.}\ }\textbf {\bibinfo
  {volume} {106}},\ \bibinfo {pages} {096401} (\bibinfo {year}
  {2011})}\BibitemShut {NoStop}%
\bibitem [{\citenamefont {Fischer}(2013)}]{fischer2013NJP}%
  \BibitemOpen
  \bibfield  {author} {\bibinfo {author} {\bibfnamefont {M.~H.}\ \bibnamefont
  {Fischer}},\ }\bibfield  {title} {\bibinfo {title} {Gap symmetry and
  stability analysis in the multi-orbital {F}e-based superconductors},\ }\href
  {https://doi.org/10.1088/1367-2630/15/7/073006} {\bibfield  {journal}
  {\bibinfo  {journal} {New J. Phys.}\ }\textbf {\bibinfo {volume} {15}},\
  \bibinfo {pages} {073006} (\bibinfo {year} {2013})}\BibitemShut {NoStop}%
\bibitem [{\citenamefont {Ramires}\ and\ \citenamefont
  {Sigrist}(2016)}]{Ramires2016PRB}%
  \BibitemOpen
  \bibfield  {author} {\bibinfo {author} {\bibfnamefont {A.}~\bibnamefont
  {Ramires}}\ and\ \bibinfo {author} {\bibfnamefont {M.}~\bibnamefont
  {Sigrist}},\ }\bibfield  {title} {\bibinfo {title} {Identifying detrimental
  effects for multiorbital superconductivity: Application to
  {${\mathrm{Sr}}_{2}{\mathrm{RuO}}_{4}$}},\ }\href
  {https://doi.org/10.1103/PhysRevB.94.104501} {\bibfield  {journal} {\bibinfo
  {journal} {Phys. Rev. B}\ }\textbf {\bibinfo {volume} {94}},\ \bibinfo
  {pages} {104501} (\bibinfo {year} {2016})}\BibitemShut {NoStop}%
\bibitem [{\citenamefont {Ramires}\ \emph {et~al.}(2018)\citenamefont
  {Ramires}, \citenamefont {Agterberg},\ and\ \citenamefont
  {Sigrist}}]{Ramires2018PRB}%
  \BibitemOpen
  \bibfield  {author} {\bibinfo {author} {\bibfnamefont {A.}~\bibnamefont
  {Ramires}}, \bibinfo {author} {\bibfnamefont {D.~F.}\ \bibnamefont
  {Agterberg}},\ and\ \bibinfo {author} {\bibfnamefont {M.}~\bibnamefont
  {Sigrist}},\ }\bibfield  {title} {\bibinfo {title} {Tailoring ${T}_{c}$ by
  symmetry principles: The concept of superconducting fitness},\ }\href
  {https://doi.org/10.1103/PhysRevB.98.024501} {\bibfield  {journal} {\bibinfo
  {journal} {Phys. Rev. B}\ }\textbf {\bibinfo {volume} {98}},\ \bibinfo
  {pages} {024501} (\bibinfo {year} {2018})}\BibitemShut {NoStop}%
\bibitem [{\citenamefont {Huang}\ \emph {et~al.}(2019)\citenamefont {Huang},
  \citenamefont {Zhou},\ and\ \citenamefont {Yao}}]{Huang2019PRB}%
  \BibitemOpen
  \bibfield  {author} {\bibinfo {author} {\bibfnamefont {W.}~\bibnamefont
  {Huang}}, \bibinfo {author} {\bibfnamefont {Y.}~\bibnamefont {Zhou}},\ and\
  \bibinfo {author} {\bibfnamefont {H.}~\bibnamefont {Yao}},\ }\bibfield
  {title} {\bibinfo {title} {Exotic {C}ooper pairing in multiorbital models of
  {${\mathrm{Sr}}_{2}{\mathrm{RuO}}_{4}$}},\ }\href
  {https://doi.org/10.1103/PhysRevB.100.134506} {\bibfield  {journal} {\bibinfo
   {journal} {Phys. Rev. B}\ }\textbf {\bibinfo {volume} {100}},\ \bibinfo
  {pages} {134506} (\bibinfo {year} {2019})}\BibitemShut {NoStop}%
\bibitem [{\citenamefont {Mackenzie}\ \emph {et~al.}(1996)\citenamefont
  {Mackenzie}, \citenamefont {Julian}, \citenamefont {Diver}, \citenamefont
  {McMullan}, \citenamefont {Ray}, \citenamefont {Lonzarich}, \citenamefont
  {Maeno}, \citenamefont {Nishizaki},\ and\ \citenamefont
  {Fujita}}]{Mackenzie1996PRL}%
  \BibitemOpen
  \bibfield  {author} {\bibinfo {author} {\bibfnamefont {A.~P.}\ \bibnamefont
  {Mackenzie}}, \bibinfo {author} {\bibfnamefont {S.~R.}\ \bibnamefont
  {Julian}}, \bibinfo {author} {\bibfnamefont {A.~J.}\ \bibnamefont {Diver}},
  \bibinfo {author} {\bibfnamefont {G.~J.}\ \bibnamefont {McMullan}}, \bibinfo
  {author} {\bibfnamefont {M.~P.}\ \bibnamefont {Ray}}, \bibinfo {author}
  {\bibfnamefont {G.~G.}\ \bibnamefont {Lonzarich}}, \bibinfo {author}
  {\bibfnamefont {Y.}~\bibnamefont {Maeno}}, \bibinfo {author} {\bibfnamefont
  {S.}~\bibnamefont {Nishizaki}},\ and\ \bibinfo {author} {\bibfnamefont
  {T.}~\bibnamefont {Fujita}},\ }\bibfield  {title} {\bibinfo {title} {Quantum
  {O}scillations in the {L}ayered {P}erovskite {S}uperconductor
  {S${\mathrm{r}}_{2}$Ru${\mathrm{O}}_{4}$}},\ }\href
  {https://doi.org/10.1103/PhysRevLett.76.3786} {\bibfield  {journal} {\bibinfo
   {journal} {Phys. Rev. Lett.}\ }\textbf {\bibinfo {volume} {76}},\ \bibinfo
  {pages} {3786} (\bibinfo {year} {1996})}\BibitemShut {NoStop}%
\bibitem [{\citenamefont {Bergemann}\ \emph {et~al.}(2000)\citenamefont
  {Bergemann}, \citenamefont {Julian}, \citenamefont {Mackenzie}, \citenamefont
  {NishiZaki},\ and\ \citenamefont {Maeno}}]{Bergemann2000PRL}%
  \BibitemOpen
  \bibfield  {author} {\bibinfo {author} {\bibfnamefont {C.}~\bibnamefont
  {Bergemann}}, \bibinfo {author} {\bibfnamefont {S.~R.}\ \bibnamefont
  {Julian}}, \bibinfo {author} {\bibfnamefont {A.~P.}\ \bibnamefont
  {Mackenzie}}, \bibinfo {author} {\bibfnamefont {S.}~\bibnamefont
  {NishiZaki}},\ and\ \bibinfo {author} {\bibfnamefont {Y.}~\bibnamefont
  {Maeno}},\ }\bibfield  {title} {\bibinfo {title} {Detailed {T}opography of
  the {F}ermi {S}urface of {${\mathrm{Sr}}_{2}{\mathrm{RuO}}_{4}$}},\ }\href
  {https://doi.org/10.1103/PhysRevLett.84.2662} {\bibfield  {journal} {\bibinfo
   {journal} {Phys. Rev. Lett.}\ }\textbf {\bibinfo {volume} {84}},\ \bibinfo
  {pages} {2662} (\bibinfo {year} {2000})}\BibitemShut {NoStop}%
\bibitem [{\citenamefont {Damascelli}\ \emph {et~al.}(2000)\citenamefont
  {Damascelli}, \citenamefont {Lu}, \citenamefont {Shen}, \citenamefont
  {Armitage}, \citenamefont {Ronning}, \citenamefont {Feng}, \citenamefont
  {Kim}, \citenamefont {Shen}, \citenamefont {Kimura}, \citenamefont {Tokura},
  \citenamefont {Mao},\ and\ \citenamefont {Maeno}}]{Damascelli2000PRL}%
  \BibitemOpen
  \bibfield  {author} {\bibinfo {author} {\bibfnamefont {A.}~\bibnamefont
  {Damascelli}}, \bibinfo {author} {\bibfnamefont {D.~H.}\ \bibnamefont {Lu}},
  \bibinfo {author} {\bibfnamefont {K.~M.}\ \bibnamefont {Shen}}, \bibinfo
  {author} {\bibfnamefont {N.~P.}\ \bibnamefont {Armitage}}, \bibinfo {author}
  {\bibfnamefont {F.}~\bibnamefont {Ronning}}, \bibinfo {author} {\bibfnamefont
  {D.~L.}\ \bibnamefont {Feng}}, \bibinfo {author} {\bibfnamefont
  {C.}~\bibnamefont {Kim}}, \bibinfo {author} {\bibfnamefont {Z.-X.}\
  \bibnamefont {Shen}}, \bibinfo {author} {\bibfnamefont {T.}~\bibnamefont
  {Kimura}}, \bibinfo {author} {\bibfnamefont {Y.}~\bibnamefont {Tokura}},
  \bibinfo {author} {\bibfnamefont {Z.~Q.}\ \bibnamefont {Mao}},\ and\ \bibinfo
  {author} {\bibfnamefont {Y.}~\bibnamefont {Maeno}},\ }\bibfield  {title}
  {\bibinfo {title} {Fermi {S}urface, {S}urface {S}tates, and {S}urface
  {R}econstruction in {${\mathrm{Sr}}_{2}{\mathrm{RuO}}_{4}$}},\ }\href
  {https://doi.org/10.1103/PhysRevLett.85.5194} {\bibfield  {journal} {\bibinfo
   {journal} {Phys. Rev. Lett.}\ }\textbf {\bibinfo {volume} {85}},\ \bibinfo
  {pages} {5194} (\bibinfo {year} {2000})}\BibitemShut {NoStop}%
\bibitem [{\citenamefont {Oguchi}(1995)}]{oguchi1995PRB}%
  \BibitemOpen
  \bibfield  {author} {\bibinfo {author} {\bibfnamefont {T.}~\bibnamefont
  {Oguchi}},\ }\bibfield  {title} {\bibinfo {title} {Electronic band structure
  of the superconductor {${\mathrm{Sr}}_{2}{\mathrm{RuO}}_{4}$}},\ }\href
  {https://link.aps.org/doi/10.1103/PhysRevB.51.1385} {\bibfield  {journal}
  {\bibinfo  {journal} {Phys. Rev. B}\ }\textbf {\bibinfo {volume} {51}},\
  \bibinfo {pages} {1385} (\bibinfo {year} {1995})}\BibitemShut {NoStop}%
\bibitem [{\citenamefont {Kim}\ \emph {et~al.}(2018)\citenamefont {Kim},
  \citenamefont {Mravlje}, \citenamefont {Ferrero}, \citenamefont {Parcollet},\
  and\ \citenamefont {Georges}}]{Kim2018PRL}%
  \BibitemOpen
  \bibfield  {author} {\bibinfo {author} {\bibfnamefont {M.}~\bibnamefont
  {Kim}}, \bibinfo {author} {\bibfnamefont {J.}~\bibnamefont {Mravlje}},
  \bibinfo {author} {\bibfnamefont {M.}~\bibnamefont {Ferrero}}, \bibinfo
  {author} {\bibfnamefont {O.}~\bibnamefont {Parcollet}},\ and\ \bibinfo
  {author} {\bibfnamefont {A.}~\bibnamefont {Georges}},\ }\bibfield  {title}
  {\bibinfo {title} {Spin-{O}rbit {C}oupling and {E}lectronic {C}orrelations in
  {${\mathrm{Sr}}_{2}{\mathrm{RuO}}_{4}$}},\ }\href
  {https://doi.org/10.1103/PhysRevLett.120.126401} {\bibfield  {journal}
  {\bibinfo  {journal} {Phys. Rev. Lett.}\ }\textbf {\bibinfo {volume} {120}},\
  \bibinfo {pages} {126401} (\bibinfo {year} {2018})}\BibitemShut {NoStop}%
\bibitem [{\citenamefont {R\o{}ising}\ \emph {et~al.}(2019)\citenamefont
  {R\o{}ising}, \citenamefont {Scaffidi}, \citenamefont {Flicker},
  \citenamefont {Lange},\ and\ \citenamefont {Simon}}]{Roising2019prr}%
  \BibitemOpen
  \bibfield  {author} {\bibinfo {author} {\bibfnamefont {H.~S.}\ \bibnamefont
  {R\o{}ising}}, \bibinfo {author} {\bibfnamefont {T.}~\bibnamefont
  {Scaffidi}}, \bibinfo {author} {\bibfnamefont {F.}~\bibnamefont {Flicker}},
  \bibinfo {author} {\bibfnamefont {G.~F.}\ \bibnamefont {Lange}},\ and\
  \bibinfo {author} {\bibfnamefont {S.~H.}\ \bibnamefont {Simon}},\ }\bibfield
  {title} {\bibinfo {title} {Superconducting order of
  {${\mathrm{Sr}}_{2}{\mathrm{RuO}}_{4}$} from a three-dimensional microscopic
  model},\ }\href {https://doi.org/10.1103/PhysRevResearch.1.033108} {\bibfield
   {journal} {\bibinfo  {journal} {Phys. Rev. Res.}\ }\textbf {\bibinfo
  {volume} {1}},\ \bibinfo {pages} {033108} (\bibinfo {year}
  {2019})}\BibitemShut {NoStop}%
\bibitem [{\citenamefont {Luke}\ \emph {et~al.}(1998)\citenamefont {Luke},
  \citenamefont {Fudamoto}, \citenamefont {Kojima}, \citenamefont {Larkin},
  \citenamefont {Merrin}, \citenamefont {Nachumi}, \citenamefont {Uemura},
  \citenamefont {Maeno}, \citenamefont {Mao}, \citenamefont {Mori},
  \citenamefont {Nakamura},\ and\ \citenamefont {Sigrist}}]{Luke1998Nature}%
  \BibitemOpen
  \bibfield  {author} {\bibinfo {author} {\bibfnamefont {G.~M.}\ \bibnamefont
  {Luke}}, \bibinfo {author} {\bibfnamefont {Y.}~\bibnamefont {Fudamoto}},
  \bibinfo {author} {\bibfnamefont {K.~M.}\ \bibnamefont {Kojima}}, \bibinfo
  {author} {\bibfnamefont {M.~I.}\ \bibnamefont {Larkin}}, \bibinfo {author}
  {\bibfnamefont {J.}~\bibnamefont {Merrin}}, \bibinfo {author} {\bibfnamefont
  {B.}~\bibnamefont {Nachumi}}, \bibinfo {author} {\bibfnamefont {Y.~J.}\
  \bibnamefont {Uemura}}, \bibinfo {author} {\bibfnamefont {Y.}~\bibnamefont
  {Maeno}}, \bibinfo {author} {\bibfnamefont {Z.~Q.}\ \bibnamefont {Mao}},
  \bibinfo {author} {\bibfnamefont {Y.}~\bibnamefont {Mori}}, \bibinfo {author}
  {\bibfnamefont {H.}~\bibnamefont {Nakamura}},\ and\ \bibinfo {author}
  {\bibfnamefont {M.}~\bibnamefont {Sigrist}},\ }\bibfield  {title} {\bibinfo
  {title} {{Time-reversal symmetry-breaking superconductivity in
  {${\mathrm{Sr}}_{2}{\mathrm{RuO}}_{4}$}}},\ }\href
  {https://doi.org/10.1038/29038} {\bibfield  {journal} {\bibinfo  {journal}
  {Nature (London)}\ }\textbf {\bibinfo {volume} {394}},\ \bibinfo {pages}
  {558} (\bibinfo {year} {1998})}\BibitemShut {NoStop}%
\bibitem [{\citenamefont {Xia}\ \emph {et~al.}(2006)\citenamefont {Xia},
  \citenamefont {Maeno}, \citenamefont {Beyersdorf}, \citenamefont {Fejer},\
  and\ \citenamefont {Kapitulnik}}]{Xia2006prl}%
  \BibitemOpen
  \bibfield  {author} {\bibinfo {author} {\bibfnamefont {J.}~\bibnamefont
  {Xia}}, \bibinfo {author} {\bibfnamefont {Y.}~\bibnamefont {Maeno}}, \bibinfo
  {author} {\bibfnamefont {P.~T.}\ \bibnamefont {Beyersdorf}}, \bibinfo
  {author} {\bibfnamefont {M.~M.}\ \bibnamefont {Fejer}},\ and\ \bibinfo
  {author} {\bibfnamefont {A.}~\bibnamefont {Kapitulnik}},\ }\bibfield  {title}
  {\bibinfo {title} {High {R}esolution {P}olar {K}err {E}ffect {M}easurements
  of {${\mathrm{Sr}}_{2}{\mathrm{RuO}}_{4}$}: {E}vidence for {B}roken
  {T}ime-{R}eversal {S}ymmetry in the {S}uperconducting {S}tate},\ }\href
  {https://doi.org/10.1103/PhysRevLett.97.167002} {\bibfield  {journal}
  {\bibinfo  {journal} {Phys. Rev. Lett.}\ }\textbf {\bibinfo {volume} {97}},\
  \bibinfo {pages} {167002} (\bibinfo {year} {2006})}\BibitemShut {NoStop}%
\bibitem [{\citenamefont {Kidwingira}\ \emph {et~al.}(2006)\citenamefont
  {Kidwingira}, \citenamefont {Strand}, \citenamefont {Van~Harlingen},\ and\
  \citenamefont {Maeno}}]{Kidwingira2006science}%
  \BibitemOpen
  \bibfield  {author} {\bibinfo {author} {\bibfnamefont {F.}~\bibnamefont
  {Kidwingira}}, \bibinfo {author} {\bibfnamefont {J.~D.}\ \bibnamefont
  {Strand}}, \bibinfo {author} {\bibfnamefont {D.~J.}\ \bibnamefont
  {Van~Harlingen}},\ and\ \bibinfo {author} {\bibfnamefont {Y.}~\bibnamefont
  {Maeno}},\ }\bibfield  {title} {\bibinfo {title} {Dynamical superconducting
  order parameter domains in {${\mathrm{Sr}}_{2}{\mathrm{RuO}}_{4}$}},\ }\href
  {https://doi.org/10.1126/science.1133239} {\bibfield  {journal} {\bibinfo
  {journal} {Science}\ }\textbf {\bibinfo {volume} {314}},\ \bibinfo {pages}
  {1267} (\bibinfo {year} {2006})}\BibitemShut {NoStop}%
\bibitem [{\citenamefont {Grinenko}\ \emph {et~al.}()\citenamefont {Grinenko},
  \citenamefont {Ghosh}, \citenamefont {Sarkar}, \citenamefont {Orain},
  \citenamefont {Nikitin}, \citenamefont {Elender}, \citenamefont {Das},
  \citenamefont {Guguchia}, \citenamefont {Brückner}, \citenamefont {Barber},
  \citenamefont {Park}, \citenamefont {Kikugawa}, \citenamefont {Sokolov},
  \citenamefont {Bobowski}, \citenamefont {Miyoshi}, \citenamefont {Maeno},
  \citenamefont {Mackenzie}, \citenamefont {Luetkens}, \citenamefont {Hicks},\
  and\ \citenamefont {Klauss}}]{Grinenko2020}%
  \BibitemOpen
  \bibfield  {author} {\bibinfo {author} {\bibfnamefont {V.}~\bibnamefont
  {Grinenko}}, \bibinfo {author} {\bibfnamefont {S.}~\bibnamefont {Ghosh}},
  \bibinfo {author} {\bibfnamefont {R.}~\bibnamefont {Sarkar}}, \bibinfo
  {author} {\bibfnamefont {J.-C.}\ \bibnamefont {Orain}}, \bibinfo {author}
  {\bibfnamefont {A.}~\bibnamefont {Nikitin}}, \bibinfo {author} {\bibfnamefont
  {M.}~\bibnamefont {Elender}}, \bibinfo {author} {\bibfnamefont
  {D.}~\bibnamefont {Das}}, \bibinfo {author} {\bibfnamefont {Z.}~\bibnamefont
  {Guguchia}}, \bibinfo {author} {\bibfnamefont {F.}~\bibnamefont {Brückner}},
  \bibinfo {author} {\bibfnamefont {M.~E.}\ \bibnamefont {Barber}}, \bibinfo
  {author} {\bibfnamefont {J.}~\bibnamefont {Park}}, \bibinfo {author}
  {\bibfnamefont {N.}~\bibnamefont {Kikugawa}}, \bibinfo {author}
  {\bibfnamefont {D.~A.}\ \bibnamefont {Sokolov}}, \bibinfo {author}
  {\bibfnamefont {J.~S.}\ \bibnamefont {Bobowski}}, \bibinfo {author}
  {\bibfnamefont {T.}~\bibnamefont {Miyoshi}}, \bibinfo {author} {\bibfnamefont
  {Y.}~\bibnamefont {Maeno}}, \bibinfo {author} {\bibfnamefont {A.~P.}\
  \bibnamefont {Mackenzie}}, \bibinfo {author} {\bibfnamefont {H.}~\bibnamefont
  {Luetkens}}, \bibinfo {author} {\bibfnamefont {C.~W.}\ \bibnamefont
  {Hicks}},\ and\ \bibinfo {author} {\bibfnamefont {H.-H.}\ \bibnamefont
  {Klauss}},\ }\href@noop {} {\bibinfo {title} {Split superconducting and
  time-reversal symmetry-breaking transitions, and magnetic order in
  {${\mathrm{Sr}}_{2}{\mathrm{RuO}}_{4}$} under uniaxial stress}},\ \Eprint
  {https://arxiv.org/abs/2001.08152} {arXiv:2001.08152 [cond-mat.supr-con]}
  \BibitemShut {NoStop}%
\bibitem [{\citenamefont {Ghosh}\ \emph {et~al.}(2020)\citenamefont {Ghosh},
  \citenamefont {Shekhter}, \citenamefont {Jerzembeck}, \citenamefont
  {Kikugawa}, \citenamefont {Sokolov}, \citenamefont {Brando}, \citenamefont
  {Mackenzie}, \citenamefont {Hicks},\ and\ \citenamefont
  {Ramshaw}}]{Ghosh2020}%
  \BibitemOpen
  \bibfield  {author} {\bibinfo {author} {\bibfnamefont {S.}~\bibnamefont
  {Ghosh}}, \bibinfo {author} {\bibfnamefont {A.}~\bibnamefont {Shekhter}},
  \bibinfo {author} {\bibfnamefont {F.}~\bibnamefont {Jerzembeck}}, \bibinfo
  {author} {\bibfnamefont {N.}~\bibnamefont {Kikugawa}}, \bibinfo {author}
  {\bibfnamefont {D.~A.}\ \bibnamefont {Sokolov}}, \bibinfo {author}
  {\bibfnamefont {M.}~\bibnamefont {Brando}}, \bibinfo {author} {\bibfnamefont
  {A.~P.}\ \bibnamefont {Mackenzie}}, \bibinfo {author} {\bibfnamefont {C.~W.}\
  \bibnamefont {Hicks}},\ and\ \bibinfo {author} {\bibfnamefont {B.~J.}\
  \bibnamefont {Ramshaw}},\ }\bibfield  {title} {\bibinfo {title}
  {Thermodynamic evidence for a two-component superconducting order parameter
  in {${\mathrm{Sr}}_{2}{\mathrm{RuO}}_{4}$}},\ }\href
  {https://doi.org/10.1038/s41567-020-1032-4} {\bibfield  {journal} {\bibinfo
  {journal} {Nat. Phys.}\ } (\bibinfo {year} {2020})},\ \bibinfo {note}
  {\href{https://doi.org/10.1038/s41567-020-1032-4}{doi:
  10.1038/s41567-020-1032-4}}\BibitemShut {NoStop}%
\bibitem [{\citenamefont {Benhabib}\ \emph {et~al.}(2020)\citenamefont
  {Benhabib}, \citenamefont {Lupien}, \citenamefont {Paul}, \citenamefont
  {Berges}, \citenamefont {Dion}, \citenamefont {Nardone}, \citenamefont
  {Zitouni}, \citenamefont {Mao}, \citenamefont {Maeno}, \citenamefont
  {Georges} \emph {et~al.}}]{Benhabib2020}%
  \BibitemOpen
  \bibfield  {author} {\bibinfo {author} {\bibfnamefont {S.}~\bibnamefont
  {Benhabib}}, \bibinfo {author} {\bibfnamefont {C.}~\bibnamefont {Lupien}},
  \bibinfo {author} {\bibfnamefont {I.}~\bibnamefont {Paul}}, \bibinfo {author}
  {\bibfnamefont {L.}~\bibnamefont {Berges}}, \bibinfo {author} {\bibfnamefont
  {M.}~\bibnamefont {Dion}}, \bibinfo {author} {\bibfnamefont {M.}~\bibnamefont
  {Nardone}}, \bibinfo {author} {\bibfnamefont {A.}~\bibnamefont {Zitouni}},
  \bibinfo {author} {\bibfnamefont {Z.~Q.}\ \bibnamefont {Mao}}, \bibinfo
  {author} {\bibfnamefont {Y.}~\bibnamefont {Maeno}}, \bibinfo {author}
  {\bibfnamefont {A.}~\bibnamefont {Georges}}, \emph {et~al.},\ }\bibfield
  {title} {\bibinfo {title} {Ultrasound evidence for a two-component
  superconducting order parameter in {${\mathrm{Sr}}_{2}{\mathrm{RuO}}_{4}$}},\
  }\href {https://doi.org/10.1038/s41567-020-01090-2} {\bibfield  {journal}
  {\bibinfo  {journal} {Nat. Phys.}\ } (\bibinfo {year} {2020})},\ \bibinfo
  {note} {\href{https://doi.org/10.1038/s41567-020-01090-2}{doi:
  10.1038/s41567-020-01090-2}}\BibitemShut {NoStop}%
\bibitem [{\citenamefont {Pustogow}\ \emph {et~al.}(2019)\citenamefont
  {Pustogow}, \citenamefont {Luo}, \citenamefont {Chronister}, \citenamefont
  {Su}, \citenamefont {Sokolov}, \citenamefont {Jerzembeck}, \citenamefont
  {Mackenzie}, \citenamefont {Hicks}, \citenamefont {Kikugawa}, \citenamefont
  {Raghu} \emph {et~al.}}]{Pustogow2019Nature}%
  \BibitemOpen
  \bibfield  {author} {\bibinfo {author} {\bibfnamefont {A.}~\bibnamefont
  {Pustogow}}, \bibinfo {author} {\bibfnamefont {Y.}~\bibnamefont {Luo}},
  \bibinfo {author} {\bibfnamefont {A.}~\bibnamefont {Chronister}}, \bibinfo
  {author} {\bibfnamefont {Y.-S.}\ \bibnamefont {Su}}, \bibinfo {author}
  {\bibfnamefont {D.~A.}\ \bibnamefont {Sokolov}}, \bibinfo {author}
  {\bibfnamefont {F.}~\bibnamefont {Jerzembeck}}, \bibinfo {author}
  {\bibfnamefont {A.~P.}\ \bibnamefont {Mackenzie}}, \bibinfo {author}
  {\bibfnamefont {C.~W.}\ \bibnamefont {Hicks}}, \bibinfo {author}
  {\bibfnamefont {N.}~\bibnamefont {Kikugawa}}, \bibinfo {author}
  {\bibfnamefont {S.}~\bibnamefont {Raghu}}, \emph {et~al.},\ }\bibfield
  {title} {\bibinfo {title} {Constraints on the superconducting order parameter
  in {${\mathrm{Sr}}_{2}{\mathrm{RuO}}_{4}$} from oxygen-17 nuclear magnetic
  resonance},\ }\href {https://doi.org/10.1038/s41586-019-1596-2} {\bibfield
  {journal} {\bibinfo  {journal} {Nature (London)}\ }\textbf {\bibinfo {volume}
  {574}},\ \bibinfo {pages} {72} (\bibinfo {year} {2019})}\BibitemShut
  {NoStop}%
\bibitem [{\citenamefont {Ishida}\ \emph {et~al.}(2020)\citenamefont {Ishida},
  \citenamefont {Manago}, \citenamefont {Kinjo},\ and\ \citenamefont
  {Maeno}}]{Ishida2019}%
  \BibitemOpen
  \bibfield  {author} {\bibinfo {author} {\bibfnamefont {K.}~\bibnamefont
  {Ishida}}, \bibinfo {author} {\bibfnamefont {M.}~\bibnamefont {Manago}},
  \bibinfo {author} {\bibfnamefont {K.}~\bibnamefont {Kinjo}},\ and\ \bibinfo
  {author} {\bibfnamefont {Y.}~\bibnamefont {Maeno}},\ }\bibfield  {title}
  {\bibinfo {title} {Reduction of the ${}^{17}${O} {K}night shift in the
  superconducting state and the heat-up effect by {NMR} pulses on
  {${\mathrm{Sr}}_{2}{\mathrm{RuO}}_{4}$}},\ }\href
  {https://doi.org/10.7566/JPSJ.89.034712} {\bibfield  {journal} {\bibinfo
  {journal} {J. Phys. Soc. Jpn.}\ }\textbf {\bibinfo {volume} {89}},\ \bibinfo
  {pages} {034712} (\bibinfo {year} {2020})}\BibitemShut {NoStop}%
\bibitem [{\citenamefont {\ifmmode \check{Z}\else
  \v{Z}\fi{}uti\ifmmode~\acute{c}\else \'{c}\fi{}}\ and\ \citenamefont
  {Mazin}(2005)}]{Zutic2005PRL}%
  \BibitemOpen
  \bibfield  {author} {\bibinfo {author} {\bibfnamefont {I.}~\bibnamefont
  {\ifmmode \check{Z}\else \v{Z}\fi{}uti\ifmmode~\acute{c}\else \'{c}\fi{}}}\
  and\ \bibinfo {author} {\bibfnamefont {I.}~\bibnamefont {Mazin}},\ }\bibfield
   {title} {\bibinfo {title} {Phase-{S}ensitive {T}ests of the {P}airing
  {S}tate {S}ymmetry in {${\mathrm{Sr}}_{2}{\mathrm{RuO}}_{4}$}},\ }\href
  {https://doi.org/10.1103/PhysRevLett.95.217004} {\bibfield  {journal}
  {\bibinfo  {journal} {Phys. Rev. Lett.}\ }\textbf {\bibinfo {volume} {95}},\
  \bibinfo {pages} {217004} (\bibinfo {year} {2005})}\BibitemShut {NoStop}%
\bibitem [{\citenamefont {Kivelson}\ \emph {et~al.}(2020)\citenamefont
  {Kivelson}, \citenamefont {Yuan}, \citenamefont {Ramshaw},\ and\
  \citenamefont {Thomale}}]{Kivelson2020npj}%
  \BibitemOpen
  \bibfield  {author} {\bibinfo {author} {\bibfnamefont {S.~A.}\ \bibnamefont
  {Kivelson}}, \bibinfo {author} {\bibfnamefont {A.~C.}\ \bibnamefont {Yuan}},
  \bibinfo {author} {\bibfnamefont {B.}~\bibnamefont {Ramshaw}},\ and\ \bibinfo
  {author} {\bibfnamefont {R.}~\bibnamefont {Thomale}},\ }\bibfield  {title}
  {\bibinfo {title} {{A proposal for reconciling diverse experiments on the
  superconducting state in {${\mathrm{Sr}}_{2}{\mathrm{RuO}}_{4}$}}},\ }\href
  {https://doi.org/10.1038/s41535-020-0245-1} {\bibfield  {journal} {\bibinfo
  {journal} {npj Quantum Mater.}\ }\textbf {\bibinfo {volume} {5}},\ \bibinfo
  {pages} {43} (\bibinfo {year} {2020})}\BibitemShut {NoStop}%
\bibitem [{\citenamefont {Yu}\ \emph {et~al.}(2018)\citenamefont {Yu},
  \citenamefont {Cheung}, \citenamefont {Raghu},\ and\ \citenamefont
  {Agterberg}}]{Yu2018PRB}%
  \BibitemOpen
  \bibfield  {author} {\bibinfo {author} {\bibfnamefont {Y.}~\bibnamefont
  {Yu}}, \bibinfo {author} {\bibfnamefont {A.~K.~C.}\ \bibnamefont {Cheung}},
  \bibinfo {author} {\bibfnamefont {S.}~\bibnamefont {Raghu}},\ and\ \bibinfo
  {author} {\bibfnamefont {D.~F.}\ \bibnamefont {Agterberg}},\ }\bibfield
  {title} {\bibinfo {title} {Residual spin susceptibility in the spin-triplet
  orbital-singlet model},\ }\href {https://doi.org/10.1103/PhysRevB.98.184507}
  {\bibfield  {journal} {\bibinfo  {journal} {Phys. Rev. B}\ }\textbf {\bibinfo
  {volume} {98}},\ \bibinfo {pages} {184507} (\bibinfo {year}
  {2018})}\BibitemShut {NoStop}%
\bibitem [{\citenamefont {Li}\ \emph {et~al.}()\citenamefont {Li},
  \citenamefont {Kikugawa}, \citenamefont {Sokolov}, \citenamefont
  {Jerzembeck}, \citenamefont {Gibbs}, \citenamefont {Maeno}, \citenamefont
  {Hicks}, \citenamefont {Nicklas},\ and\ \citenamefont {Mackenzie}}]{Li2019}%
  \BibitemOpen
  \bibfield  {author} {\bibinfo {author} {\bibfnamefont {Y.~S.}\ \bibnamefont
  {Li}}, \bibinfo {author} {\bibfnamefont {N.}~\bibnamefont {Kikugawa}},
  \bibinfo {author} {\bibfnamefont {D.~A.}\ \bibnamefont {Sokolov}}, \bibinfo
  {author} {\bibfnamefont {F.}~\bibnamefont {Jerzembeck}}, \bibinfo {author}
  {\bibfnamefont {A.~S.}\ \bibnamefont {Gibbs}}, \bibinfo {author}
  {\bibfnamefont {Y.}~\bibnamefont {Maeno}}, \bibinfo {author} {\bibfnamefont
  {C.~W.}\ \bibnamefont {Hicks}}, \bibinfo {author} {\bibfnamefont
  {M.}~\bibnamefont {Nicklas}},\ and\ \bibinfo {author} {\bibfnamefont {A.~P.}\
  \bibnamefont {Mackenzie}},\ }\href@noop {} {\bibinfo {title} {High
  sensitivity heat capacity measurements on
  {${\mathrm{Sr}}_{2}{\mathrm{RuO}}_{4}$} under uniaxial pressure}},\ \Eprint
  {https://arxiv.org/abs/1906.07597} {arXiv:1906.07597 [cond-mat.supr-con]}
  \BibitemShut {NoStop}%
\bibitem [{\citenamefont {NishiZaki}\ \emph {et~al.}(2000)\citenamefont
  {NishiZaki}, \citenamefont {Maeno},\ and\ \citenamefont
  {Mao}}]{NishiZaki2000JPS}%
  \BibitemOpen
  \bibfield  {author} {\bibinfo {author} {\bibfnamefont {S.}~\bibnamefont
  {NishiZaki}}, \bibinfo {author} {\bibfnamefont {Y.}~\bibnamefont {Maeno}},\
  and\ \bibinfo {author} {\bibfnamefont {Z.}~\bibnamefont {Mao}},\ }\bibfield
  {title} {\bibinfo {title} {Changes in the superconducting state of
  {${\mathrm{Sr}}_{2}{\mathrm{RuO}}_{4}$} under magnetic fields probed by
  specific heat},\ }\href {https://doi.org/10.1143/JPSJ.69.572} {\bibfield
  {journal} {\bibinfo  {journal} {J. Phys. Soc. Jpn.}\ }\textbf {\bibinfo
  {volume} {69}},\ \bibinfo {pages} {572} (\bibinfo {year} {2000})}\BibitemShut
  {NoStop}%
\bibitem [{\citenamefont {Bonalde}\ \emph {et~al.}(2000)\citenamefont
  {Bonalde}, \citenamefont {Yanoff}, \citenamefont {Salamon}, \citenamefont
  {Van~Harlingen}, \citenamefont {Chia}, \citenamefont {Mao},\ and\
  \citenamefont {Maeno}}]{Bonalde2000PRL}%
  \BibitemOpen
  \bibfield  {author} {\bibinfo {author} {\bibfnamefont {I.}~\bibnamefont
  {Bonalde}}, \bibinfo {author} {\bibfnamefont {B.~D.}\ \bibnamefont {Yanoff}},
  \bibinfo {author} {\bibfnamefont {M.~B.}\ \bibnamefont {Salamon}}, \bibinfo
  {author} {\bibfnamefont {D.~J.}\ \bibnamefont {Van~Harlingen}}, \bibinfo
  {author} {\bibfnamefont {E.~M.~E.}\ \bibnamefont {Chia}}, \bibinfo {author}
  {\bibfnamefont {Z.~Q.}\ \bibnamefont {Mao}},\ and\ \bibinfo {author}
  {\bibfnamefont {Y.}~\bibnamefont {Maeno}},\ }\bibfield  {title} {\bibinfo
  {title} {Temperature {D}ependence of the {P}enetration {D}epth in
  {${\mathrm{Sr}}_{2}{\mathrm{RuO}}_{4}$}: Evidence for {N}odes in the {G}ap
  {F}unction},\ }\href {https://doi.org/10.1103/PhysRevLett.85.4775} {\bibfield
   {journal} {\bibinfo  {journal} {Phys. Rev. Lett.}\ }\textbf {\bibinfo
  {volume} {85}},\ \bibinfo {pages} {4775} (\bibinfo {year}
  {2000})}\BibitemShut {NoStop}%
\bibitem [{\citenamefont {Lupien}\ \emph {et~al.}(2001)\citenamefont {Lupien},
  \citenamefont {MacFarlane}, \citenamefont {Proust}, \citenamefont
  {Taillefer}, \citenamefont {Mao},\ and\ \citenamefont
  {Maeno}}]{Lupien2001PRL}%
  \BibitemOpen
  \bibfield  {author} {\bibinfo {author} {\bibfnamefont {C.}~\bibnamefont
  {Lupien}}, \bibinfo {author} {\bibfnamefont {W.~A.}\ \bibnamefont
  {MacFarlane}}, \bibinfo {author} {\bibfnamefont {C.}~\bibnamefont {Proust}},
  \bibinfo {author} {\bibfnamefont {L.}~\bibnamefont {Taillefer}}, \bibinfo
  {author} {\bibfnamefont {Z.~Q.}\ \bibnamefont {Mao}},\ and\ \bibinfo {author}
  {\bibfnamefont {Y.}~\bibnamefont {Maeno}},\ }\bibfield  {title} {\bibinfo
  {title} {Ultrasound {A}ttenuation in {${\mathrm{Sr}}_{2}{\mathrm{RuO}}_{4}$}:
  An {A}ngle-{R}esolved {S}tudy of the {S}uperconducting {G}ap {F}unction},\
  }\href {https://doi.org/10.1103/PhysRevLett.86.5986} {\bibfield  {journal}
  {\bibinfo  {journal} {Phys. Rev. Lett.}\ }\textbf {\bibinfo {volume} {86}},\
  \bibinfo {pages} {5986} (\bibinfo {year} {2001})}\BibitemShut {NoStop}%
\bibitem [{\citenamefont {Firmo}\ \emph {et~al.}(2013)\citenamefont {Firmo},
  \citenamefont {Lederer}, \citenamefont {Lupien}, \citenamefont {Mackenzie},
  \citenamefont {Davis},\ and\ \citenamefont {Kivelson}}]{Firmo2013PRB}%
  \BibitemOpen
  \bibfield  {author} {\bibinfo {author} {\bibfnamefont {I.~A.}\ \bibnamefont
  {Firmo}}, \bibinfo {author} {\bibfnamefont {S.}~\bibnamefont {Lederer}},
  \bibinfo {author} {\bibfnamefont {C.}~\bibnamefont {Lupien}}, \bibinfo
  {author} {\bibfnamefont {A.~P.}\ \bibnamefont {Mackenzie}}, \bibinfo {author}
  {\bibfnamefont {J.~C.}\ \bibnamefont {Davis}},\ and\ \bibinfo {author}
  {\bibfnamefont {S.~A.}\ \bibnamefont {Kivelson}},\ }\bibfield  {title}
  {\bibinfo {title} {Evidence from tunneling spectroscopy for a
  quasi-one-dimensional origin of superconductivity in
  {${\mathrm{Sr}}_{2}{\mathrm{RuO}}_{4}$}},\ }\href
  {https://doi.org/10.1103/PhysRevB.88.134521} {\bibfield  {journal} {\bibinfo
  {journal} {Phys. Rev. B}\ }\textbf {\bibinfo {volume} {88}},\ \bibinfo
  {pages} {134521} (\bibinfo {year} {2013})}\BibitemShut {NoStop}%
\bibitem [{\citenamefont {Hassinger}\ \emph {et~al.}(2017)\citenamefont
  {Hassinger}, \citenamefont {Bourgeois-Hope}, \citenamefont {Taniguchi},
  \citenamefont {Ren\'e~de Cotret}, \citenamefont {Grissonnanche},
  \citenamefont {Anwar}, \citenamefont {Maeno}, \citenamefont
  {Doiron-Leyraud},\ and\ \citenamefont {Taillefer}}]{Hassinger2017PRX}%
  \BibitemOpen
  \bibfield  {author} {\bibinfo {author} {\bibfnamefont {E.}~\bibnamefont
  {Hassinger}}, \bibinfo {author} {\bibfnamefont {P.}~\bibnamefont
  {Bourgeois-Hope}}, \bibinfo {author} {\bibfnamefont {H.}~\bibnamefont
  {Taniguchi}}, \bibinfo {author} {\bibfnamefont {S.}~\bibnamefont {Ren\'e~de
  Cotret}}, \bibinfo {author} {\bibfnamefont {G.}~\bibnamefont
  {Grissonnanche}}, \bibinfo {author} {\bibfnamefont {M.~S.}\ \bibnamefont
  {Anwar}}, \bibinfo {author} {\bibfnamefont {Y.}~\bibnamefont {Maeno}},
  \bibinfo {author} {\bibfnamefont {N.}~\bibnamefont {Doiron-Leyraud}},\ and\
  \bibinfo {author} {\bibfnamefont {L.}~\bibnamefont {Taillefer}},\ }\bibfield
  {title} {\bibinfo {title} {Vertical {L}ine {N}odes in the {S}uperconducting
  {G}ap {S}tructure of {${\mathrm{Sr}}_{2}{\mathrm{RuO}}_{4}$}},\ }\href
  {https://doi.org/10.1103/PhysRevX.7.011032} {\bibfield  {journal} {\bibinfo
  {journal} {Phys. Rev. X}\ }\textbf {\bibinfo {volume} {7}},\ \bibinfo {pages}
  {011032} (\bibinfo {year} {2017})}\BibitemShut {NoStop}%
\bibitem [{\citenamefont {Sharma}\ \emph {et~al.}(2020)\citenamefont {Sharma},
  \citenamefont {Edkins}, \citenamefont {Wang}, \citenamefont {Kostin},
  \citenamefont {Sow}, \citenamefont {Maeno}, \citenamefont {Mackenzie},
  \citenamefont {Davis},\ and\ \citenamefont {Madhavan}}]{sharma2020PNAS}%
  \BibitemOpen
  \bibfield  {author} {\bibinfo {author} {\bibfnamefont {R.}~\bibnamefont
  {Sharma}}, \bibinfo {author} {\bibfnamefont {S.~D.}\ \bibnamefont {Edkins}},
  \bibinfo {author} {\bibfnamefont {Z.}~\bibnamefont {Wang}}, \bibinfo {author}
  {\bibfnamefont {A.}~\bibnamefont {Kostin}}, \bibinfo {author} {\bibfnamefont
  {C.}~\bibnamefont {Sow}}, \bibinfo {author} {\bibfnamefont {Y.}~\bibnamefont
  {Maeno}}, \bibinfo {author} {\bibfnamefont {A.~P.}\ \bibnamefont
  {Mackenzie}}, \bibinfo {author} {\bibfnamefont {J.~S.}\ \bibnamefont
  {Davis}},\ and\ \bibinfo {author} {\bibfnamefont {V.}~\bibnamefont
  {Madhavan}},\ }\bibfield  {title} {\bibinfo {title} {Momentum-resolved
  superconducting energy gaps of {${\mathrm{Sr}}_{2}{\mathrm{RuO}}_{4}$} from
  quasiparticle interference imaging},\ }\href
  {https://doi.org/10.1073/pnas.1916463117} {\bibfield  {journal} {\bibinfo
  {journal} {Proc. Natl. Acad. Sci. (USA)}\ }\textbf {\bibinfo {volume}
  {117}},\ \bibinfo {pages} {5222} (\bibinfo {year} {2020})}\BibitemShut
  {NoStop}%
\end{thebibliography}%

\end{document}